\documentclass[12pt,preprint]{aastex}
\usepackage{natbib}
\usepackage{psfig}
\usepackage{epsfig}
\usepackage{graphicx}
\title{`Missing Link' Clouds in the Southern Galactic Plane Survey}
\author{D. W. Kavars}
\affil{Department of Astronomy, University of Minnesota, 116 Church St. SE, Minneapolis, MN, 55455, dkavars@astro.umn.edu}
\author{J. M. Dickey}
\affil{Department of Astronomy, University of Minnesota, 116 Church St. SE, Minneapolis, MN, 55455}
\affil{School of Mathematics and Physics, University of Tasmania,  GPO Box 252-21, Hobart, Tasmania, 7001, john.dickey@utas.edu.au }
\author{N. M. McClure-Griffiths}
\affil{Australia Telescope National Facility, CSIRO, PO Box 76, Epping NSW 1710, Australia, Naomi.McClure-Griffiths@atnf.csiro.au}
\author{B. M. Gaensler}
\affil{Harvard-Smithsonian Center for Astrophysics, 60 Garden St. MS-6, Cambridge, MA, 02138, bgaensler@cfa.harvard.edu}
\author{A. J. Green}
\affil{Astrophysics Department, School of Physics, University of Sydney, Sydney, NSW 2006, Australia, agreen@physics.usyd.edu.au}

\begin{abstract}
We present an automated routine to search for HI self-absorption features within the Southern Galactic Plane Survey (SGPS).  The data were taken with the Australia Telescope Compact Array (ATCA) and the Parkes Radio Telescope and encompass 3$^{\circ}$ $\times$ 105$^{\circ}$ of sky in the Galactic plane.  We apply our routine to this entire region and derive spin temperatures and column densities for 70 of the larger HISA complexes, finding spin temperatures ranging from 6-41 K with HI number densities of a few  cm$^{-3}$.  These `missing link' clouds fill in the spin temperature and density gaps between dense molecular clouds and diffuse atomic clouds.  We compare the HI emission with $^{12}$CO emission and find that $\sim$60\% of detected HI self-absorption is correlated in space and in velocity with a molecular counterpart.  This is potentially due to a molecular/atomic gas transition.  We also compare HI self-absorption with Galactic spiral arms and discuss the possibility of using it as a spiral arm tracer.  
\end{abstract}
\keywords{ISM: atoms --- radio lines: ISM --- ISM: clouds}

\begin{document}

\section{Introduction}

Neutral hydrogen (HI) self-absorption (HISA) was discovered only four years after the first detection of HI emission.  HISA is observed as a very narrow absorption dip of only a few (1-4) km s$^{-1}$ in HI profiles \citep{ 1974AJ.....79..527K} and occurs whenever there is a cold HI cloud in front of a warmer background at the same radial velocity.  A number of studies have looked at HISA, finding spin temperatures of 8-60 K \citep{1977A&A....54..933W, 1980ApJ...242..416L, 2000ApJ...540..851G, 2003ApJ...598.1048K}.  One of the more spectacular features is the cold, dark arc of \citet{2001Natur.412..308K}, which has a very cold spin temperature of only 10 K.  Much of the earlier work, including that of \citet{1972A&A....18...55R}, and even a more recent study by \citet{2001ApJ...555..868M}, used single dish observations only, with limited angular resolution.  Although single dish observations are successful at finding HISA, HISA features can have a very complicated morphology and are often very small on the sky with angular sizes of only a few arcminutes.  For a comprehensive view of HISA, interferometer observations, along with single dish observations, are needed.  The highest resolution single-dish surveys (3.2$^{\prime}$ FWHM for Arecibo) can also find HISA, as shown by \citep{1994ApJ...436..117K}.  The improvement in the beam width from a single dish to a combined mosaic survey has a huge effect in the ability to distinguish the smaller or more sinuous HISA structures.  In addition, the continuum subtraction can be done much more precisely in the uv-data than in the map plane.  Surveys such as the Canadian Galactic Plane Survey (CGPS) \citep{2003AJ....125.3145T}, VLA Galactic Plane Survey (VGPS) \citep{2002stdd.conf...68T}, and Southern Galactic Plane Survey (SGPS) \citep{2001ApJ...551..394M} have angular resolutions at the arcminute level and thus provide a wealth of data in which to look for and study the HISA phenomenon.  They also have a fine enough velocity sampling (0.82 km s$^{-1}$) that the narrow absorption dips in the emission profiles are easily seen.

Although recent surveys have provided a large amount of data in which to search for HISA, complications still remain.  Since HISA only exists when there is a background of warmer HI emission, not all of the cold HI clouds in the Galaxy are visible and detectable.  In order to estimate the spin temperature of a given cloud, knowledge of the off-cloud brightness temperature is needed.  Different studies apply different techniques to measure this quantity \citep{1983AJ.....88..658H, 1993A&A...276..531F, 2003ApJ...598.1048K, 2004Gibson}.  There is also the difficulty of separating the spin temperature and the optical depth, which cannot be found simultaneously.  The spin temperature cannot be found without prior knowledge of the optical depth of a given cloud or by making assumptions about the optical depth.    Both are unknowns in the radiative transfer equation, discussed in $\S$ 3, with too few constraints.  Most studies, including this one, have made various assumptions to minimize the complications in order to develop a better understanding of the role that HISA plays in the ISM.  In this paper, to compute a range of spin temperatures, we assume a value for the optical depth.  In the special case where a HISA cloud is observed to transition from self-absorption to emission, stronger constraints can be placed on the spin temperature and optical depth \citep{Kerton...2005}. 

Recent results suggest that HISA probes a population of `missing link' clouds \citep{2002ASPC..276..235G, 2003ApJ...598.1048K} that fill in the spin temperature and number density gaps between the diffuse atomic clouds seen in absorption towards continuum sources and the dense molecular clouds.  Diffuse atomic clouds are found to have spin temperatures of 25-100 K and densities of a few tens to 100 cm$^{-3}$ \citep{1988gera.book...95K, 1990ARA&A..28..215D}, while molecular clouds have kinetic temperatures of $\leq$ 10 K and densities that exceed 10$^{3}$ cm$^{-3}$. Clouds filling the spin temperature gap are seen in 21cm continuum absorption studies towards extragalactic background sources \citep{2003ApJ...586.1067H}. 

Despite the number of studies that have looked at HISA and $^{12}$CO, we still do not have a clear understanding of the relationship between the two.  It is possible that these clouds are in a transition between atomic gas and molecular gas.  Some studies have found a correlation between the atomic gas and molecular gas \citep{1979A&AS...35..129B, 1981ApJ...243..778L, 1988A&A...207..145J}, while more recent studies using data from the CGPS found no correlation in velocity or space in many examples \citep{2000ApJ...540..851G, 2001Natur.412..308K, 2002ASPC..276..235G}.  Part of the problem is a selection effect.  Studies that search for HISA by starting out with a collection of dark dust clouds or globules, such as those of \citet{1974AJ.....79..527K}, \citet{1980ApJ...242..416L}, \citet{1980ApJ...241..183B}, \citet{1988A&A...205..225S}, and \citet{2003ApJ...585..823L}, often find HI clouds with very cold temperatures of 10-15 K and smaller HI column densities than this study.  A small residual amount of HI can reside in dense molecular clouds due to cosmic-ray ionization and photodissociating UV radiation \citep{2003A&A...398..621L, 2003ApJ...585..823L}.  In such clouds there should be a clear correlation between the molecular gas and the atomic gas.  But the same can not be said for studies looking for HISA based not on a molecular tracer, but on their self-absorption signature.  HISA clouds that are associated with molecular gas may be in a transition stage if the external pressure or radiation field has changed.   This is discussed further in $\S$ 5.  

How is HISA distributed throughout the Galaxy? \citet{1987ApJ...317..646P} show that over the longitude range $l$ = 36$^{\circ}$ - 40$^{\circ}$, HISA and $^{12}$CO are well correlated within the Sagittarius spiral arm in velocity-longitude space.  Although the cold atomic hydrogen is not necessarily mixed with the molecular hydrogen it must be located nearby, possibly in a shell surrounding the molecular cloud.  However, the same study shows that in the inter-arm regions this correlation does not hold.  HISA is not found to be mixed or near a molecular component.  One possibility is that as cold HI clouds enter a spiral arm region, the change in environment and interaction with a spiral density wave converts the cloud from atomic to molecular gas.  Thus, clouds within the spiral arms show a mix of HI and H$_2$, but those outside do not.  This transformation has been proposed for the giant HI cloud studied by \citet{2001ApJ...555..868M}.

To study the role that HISA plays in the ISM we look through the entire SGPS data set, discussed in $\S$ 2, using an automated searching routine to find HISA features which we present in $\S$ 3.  In $\S$ 4 we summarize our spin temperature and density results, and in $\S$ 5 we discuss the implications of the results.

\section{Data}

The SGPS is one of three large sky surveys that image HI line emission and 21cm continuum in the plane of the Galaxy.  The other two surveys are the CGPS and the VGPS.  Together, the three make up the International Galactic Plane Survey (IGPS).  The SGPS combines interferometer data from the Australia Telescope Compact Array (ATCA) and single dish data from the Parkes 64 m telescope.  It maps the region of 253$^{\circ}$ $\leq$ $l$ $\leq$ 358$^{\circ}$ and $-$1.3$^{\circ}$ $\leq$ $b$ $\leq$ $+$1.3$^{\circ}$ with an angular resolution of 2$^{\prime}$, a channel spacing of 0.82 km s$^{-1}$ and a rms brightness sensitivity per channel of $\sim$2-4 K, depending on the brightness of the HI emission \citep{2001PhDT.........9M}.\footnote{The SGPS HI images are available at http://www.atnf.csiro.au/research/HI/sgps/queryForm.html}  The SGPS data are divided into cubes of 5$^{\circ}$ $\times$ 2.6$^{\circ}$ centered on $b$ = 0$^{\circ}$ and longitudes that are multiples of 5$^{\circ}$.  A further description of the data is detailed in \citet{SGPS}.

An example of a HISA complex seen in the SGPS is given in Figure 1.  In this image there are a number of small HISA clouds centered around $l$ = 304.2$^{\circ}$, $b$ = $-$0.1$^{\circ}$, and $v$ = $-$33.80 km s$^{-1}$.  Black contours outline the regions that pass the HISA threshold test described in $\S$ 3. A cross marks the position at which the velocity profile was taken.  The sharp drop in brightness temperature followed by a sharp rise over a span of only a few km s$^{-1}$ is a typical characteristic of what we are able to detect as HISA.

In $\S$ 5.2 we compare the HI emission from the SGPS to $^{12}$CO emission from \citet{1987ApJ...322..706D} and \citet{2001ApJ...547..792D} who used the 1.2m telescope at the Cerro Tololo Interamerican Observatory in Chile.  The $^{12}$CO data has an effective angular resolution of 0.125$^{\circ}$ and a channel spacing of 1.3 km s$^{-1}$.  The rms noise level is 0.12 K.  

\section{Search Routine}

One simple approach is to scan through the HI spectral line data cubes with the eye, looking for cloud-like features that drop abruptly in HI brightness temperature and quickly rise again in a span of a few velocity channels (a few km s$^{-1}$).  A search by eye has been done for the SGPS, finding a number of self-absorption features.  This method is good for finding the largest of the HISA features, those that are a half degree or greater in size.  However, features that may be very small or at a large distance from us and only make up a few pixels on an HI map are nearly impossible to detect by eye alone.  Thus, to avoid the bias of finding only the largest features, an objective, automated search technique is required.  An automated search of HISA has also been done for the CGPS \citep{2004GibsonA}.

The searching routine that we developed takes advantage of the shape of the velocity profile of most typical HISA clouds.  Clouds without the nominal velocity profile shape may escape detection.  This is discussed further at the end of this section.  An HI brightness temperature map of a second cloud is shown Figure 2.  The velocity profile (Figure 3) drops sharply in brightness temperature (40 K over 3 velocity channels) and quickly rises.  This sharp drop and rise is a characteristic of detectable HISA with this method, and can be quantified by the first derivative of the velocity profile.  
The first derivative, shown in Figure 3 and given by Equation 1, contains a negative peak, corresponding to the sharp temperature drop in the velocity profile, quickly followed by a positive peak, corresponding to the rise in temperature.  

\begin{equation}
D_{first}(i,j,k) = \frac{T_{B}(i,j,k+1) - T_{B}(i,j,k-1)}{V(i,j,k+1) - V(i,j,k-1)}
\end{equation}

\noindent In Equation 1, $T_{B}(i,j,k+1) - T_{B}(i,j,k-1)$ is the difference in brightness temperature over the velocity interval, $V(i,j,k+1) - V(i,j,k-1) = 1.65 \ km \ s^{-1}$, where $i$ and $j$ correspond to $l$ and $b$, respectively, and $k$ corresponds to the radial velocity, $v$.  The routine looks through the entire SGPS for all features that show this double-peaked structure in the first derivative profile.  To strengthen our search, we add a second criterion.  In the first derivative, there is a sharp change from the negative peak to the positive peak.  This change is shown by taking the second derivative of the HI velocity profile, shown in Figure 3 and given by Equation 2.  

\begin{equation}
D_{second}(i,j,k) = \frac{D_{first}(i,j,k+2) - D_{first}(i,j,k-1)}{V(i,j,k+2) - V(i,j,k-1)}
\end{equation}

\noindent The difference in the first derivative values, over the velocity interval $V(i,j,k+2) - V(i,j,k-1) = 2.46 \ km \ s^{-1}$, is given by $D_{first}(i,j,k+2) - D_{first}(i,j,k-1)$.  The result is a single, positive peak at the central velocity of the HISA cloud.  These particular constant velocity intervals were chosen such that our search routine produced the highest percentage of detectable HISA (detectable HISA vs. actual HISA) and the fewest number of false detections.  Since the first and second derivative peaks may not align perfectly in velocity we compare $D_{second}(i,j,k)$ to $D_{first}(i,j,k)$, $D_{first}(i,j,k+1)$, and $D_{first}(i,j,k-1)$.  Using this method, we can build a catalog of HISA clouds by defining HISA pixels if and only if they show the negative/positive peaks in the first derivative and the positive peak in the second derivative.

To avoid confusion with random HI emission fluctuations, thresholds are set on the first and second derivative values.  Only pixels with derivative values above a specified limit are defined as HISA pixels.  The threshold is determined by looking at the number distribution function of the values of the first and second derivatives.  This is shown in Figure 4 for the cube centered around $l$ = 330$^{\circ}$, ranging from $l$ = 328$^{\circ}$ to $l$ = 333$^{\circ}$.  On both sides, the distribution decreases as a power law, and straight lines, due to the bulk of HI structure in the Galaxy, are fit to these portions on the plot.  In addition, a second straight line can be fit with a lesser slope representing a population of HISA pixels.  The second derivative number distribution can be fit with two separate lines on the positive side of the distribution only.  On the negative side, only one line can be fit.  This positive-side tail is due to the presence of a HISA population, with a high second derivative, and is not present in cubes that contain very little HISA.  We use the point at which the tail breaks away from the random HI emission fluctuations as our threshold in our initial search for HISA.  In all cubes the second derivative threshold is set at 8 K (km s$^{-1})^{-2}$. This is the same as looking for negative to positive peak changes of [8 K (km s$^{-1})^{-2}$] $\times$ [3 (4 channels of information)] $\times$ [0.82 km s$^{-1}$] = 20 K (km s$^{-1}$)$^{-1}$, corresponding to pixels with a first derivative absolute value of 10 K (km s$^{-1}$)$^{-1}$. Only those features that show the double peaked nature with peaks of at least $\pm$10 K (km s$^{-1}$)$^{-1}$ and have corresponding second derivative peaks of at least 8 K (km s$^{-1})^{-2}$ pass the coincidental threshold derivative test and are defined as HISA.

If the second derivative peak threshold of 8 K (km s$^{-1})^{-2}$ is the lone criterion in our search, one expects non-HISA pixels to contaminate our HISA detections in the 8-12 K (km s$^{-1})^{-2}$ range.  The requirement of both the second derivative and first derivative peaks removes many of the non-HISA pixels.  A non-HISA pixel with a high second derivative value does not necessarily have both a first derivative positive peak and a first derivative negative peak above/below the threshold limits.  A spatial filter is then applied to remove stand-alone HISA pixels.  These individual pixels are not especially meaningful since they are below the angular resolution of the SGPS survey.  Only collections of adjacent pixels at the resolution of the survey are considered.  

In addition, a visual check is performed as a test of our automated search routine.  The visual check finds some dark features near detected HISA clouds, seen in Figures 1 and 2.  These dark features may be an extension of the HISA outlined from the search, escaping detection due to broader linewidths, differing values of $T_{off}$ on each side of the self-absorption dip, and less-pronounced self-absorption dips.  By lowering the derivative thresholds to 7.5 (km s$^{-1}$)$^{-1}$ and 6 (km s$^{-1})^{-2}$, most of these dark features are detected as HISA, with the total number of detected HISA pixels increasing by $\sim$25\% in Figures 1 and 2, but this greatly increases the risk of contaminating the HISA features with non-HISA features.  Thus we set our derivative thresholds high enough such that there is minimal confusion between HISA and random HI emission fluctuations, and note that HISA with the above properties could escape detection from this search.  Similar dark features that escape detection are seen under the same conditions that detected HISA is observed (i.e. with associated molecular emission, in the absence of molecular mission, near galactic spiral arms, and in inter-arm regions).  Although this limit exists in our automated HISA search method, it does not have a serious effect on discussions presented in $\S$ 5 since the appearance of the undetected dark features is random. 

It is important to mention that the above procedures could result in potential confusion between HISA and absorption due to background continuum sources.  There are a number of catalogued HII regions and supernova remnants within the SGPS \citep{1987A&A...171..261C, 1996A&AS..118..329W} as well as a number of uncatalogued continuum sources.  Any of these could produce a line profile that would pass as HISA in our search routine, but is not necessarily due to HISA.  To avoid this confusion, we remove all pixels with a continuum brightness temperature greater than 25 K, when continuum absorption becomes significant.  If a HISA cloud lies along a line of sight towards a background continuum source, it will remain undetected.  However, we can visually study these occurences for potential information \citep{2003ApJ...598.1048K}.
 
The above steps minimize the occurence of false detections, but do not eliminate them altogether.  A final visual check finds $\sim$ 5\% of detected HISA pixels are false detections.  The majority of false detections are due to HI emission gaps where the HI brightness temperature is low (T$_B$ $\leq$ 60-70 K), and several continuum sources with temperatures just under our threshold limit of 25 K.  The continuum threshold could be lowered further to reduce the percentage of false detections, but the result is an increase in the number HISA pixels that escape detection.  Many of the false detections can be flagged manually with little difficulty and any affects in the following discussions are minimal.  

\section{Results}
\subsection{Spin Temperature}

After searching for HISA clouds, we analyze the resulting features using the procedure outlined in \citet{2003ApJ...598.1048K}.  We use our four component model of the ISM (equations 2-6), consisting of foreground and background HI emission, a HISA cloud, and four sources of diffuse continuum.  Working through the radiative transfer, the spin temperature is found as a function of the optical depth, given by

\begin{equation}
T_{s,HISA} = T_{cont} + pT_{off} + \frac{T_{on}-T_{off}}{1-e^{-{\tau}_{HISA}}}\ .
\end{equation}

The observed HI brightness temperature of the cloud is $T_{on}$, the off-cloud HI brightness temperature is $T_{off}$, and $T_{cont}$ represents contributions from all sources of continuum.  To determine $T_{off}$, we interpolate the off-cloud brightness temperature in the spectral domain across the self-absorption dip by fitting a parabola to the individual velocity profiles.  This procedure is explained in further detail by \citet{2003ApJ...598.1048K}. The remaining variable, $p$ describes the fraction of HI emission that orginates behind the cloud and is given by

\begin{equation}
p = \frac{[T_s]_{bg}(1-e^{-{\tau}_{bg}})}{[T_s]_{fg}(1-e^{-{\tau}_{fg}}) + [T_s]
_{bg}(1-e^{-{\tau}_{bg}}) + T_{c}(e^{-{\tau}_{bg}}-1)}\ .
\end{equation}

$[T_s]_{fg}$ and $[T_s]_{bg}$ are the spin temperatures of the foreground and background HI emission, respectively, and $\tau_{fg}$ and $\tau_{bg}$ are the optical depths of the corresponding HI emission.  $T_c$ is the continuum emission from behind the cloud.  Although a solution for the spin temperature and optical depth can lie anywhere on the curve given by Equation 3 (see Figure 6 of \citet{2003ApJ...598.1048K}), we can immediately find the upper limit spin temperature corresponding to an infinite optical depth.  However, it is likely that the spin temperature is slightly cooler such that the optical depth is closer to unity.  This is shown by many studies, including those of \citet{2000ApJ...540..851G} and \citet{2004ApJ...603..560S}.  Also, in almost all cases, the velocity profiles are Gaussian shaped, so the optical depth is not large enough to cause an appreciable amount of saturation in the line. As a result, we cannot use saturation broadening to limit the spin temperature.

Using Equation 3 we compute spin temperatures for all HISA features found from our search routine.  Spin temperatures are computed for an infinite optical depth and a more likely optical depth of one.  It is unlikely that the optical depth is much below this value, as this pushes the spin temperature to extremely cool levels ($T_s \le$ 10 K).  The lower limit optical depth for cloud \#30 (Figure 1, Table 2) is 0.6, given by Equation 3, for $T_s$ $=$ 3 K and $p$ $=$ 0.75.  Although the optical depth could be higher than 1, it is limited by the absence of optical obscuration by dust.  For optical depths around 1, the spin temperature is weakly dependent on tau.  If the optical depth of cloud \#30 were 0.75, the spin temperature would decrease from the listed 29 K to 17 K.  For an optical depth of 2, the spin temperature increases to 51 K.  This in turn has an effect on the column density and number density of the cloud.  In this example, doubling the optical depth increases the column density and number density values by a factor of 3-4.  Spin temperature ranges for each SGPS cube are listed in Table 1.  The ranges are defined such that a quarter of the pixels lie below the lower limit and a quarter lie above the upper limit given.  A spin temperature histogram of the cube centered on $l$ = 310$^{\circ}$ is shown in Figure 5, for $\tau$ = 1.  Half of the HISA pixels lie in the spin temperature range give in Table 1, but there are many features with much cooler temperatures ($T_s$ $\le$ 20 K.).  

 In all cases, the parameter, $p$, is set to 0.75.  Thus, the majority of the HI emission lies behind the cloud, allowing any cloud to show up as a self-absorbing feature.  Although it is possible that $p$ could be closer to 0, with less gas behind the cloud than in front, it is likely this is not the case.  In the fourth quadrant of the Galaxy, a cloud at a negative velocity can lie at one of two distances from us, establishing a distance ambiquity for such clouds.  \citet{1981ApJ...246...74L} suggest that HISA can be used to solve the distance ambiquity for molecular clouds.  Near distance molecular clouds will show HISA while those at the far distance will not due to the lack of background to absorb.  This is most recently shown for the molecular cloud GRSMC 45.6+0.3 \citep{2002ApJ...566L..81J}.  Therefore it is most likely that all of the clouds found in our search are at the near kinematic distance and $p$ is closer to 1.  It has also been observed that far-distance clouds can provide their own emission backgrounds for HISA in spiral density waves \citep{2002ASPC..276..235G, 2004Gibson}.  If a given cloud is at the far kinematic distance, its spin temperature is lower than that presented here.  For example, if $p$ $=$ 0.6 for cloud \#30, the spin temperature is 10 K, assuming an optical depth of 1.  Pushing $p$ to values lower than this gives unrealistically low temperatures, providing a lower limit for $p$.  Conversely, if $p$ is set to the upper limit of 1, the spin temperature for cloud \#30 becomes 60 K.  Similar to changes in the optical depth, changes to $p$ change the column density and number density 
by up to a factor of $\sim$2.

\subsection{Distance and Size}

To explore the density range of HISA, we select the largest HISA clouds, all greater than 0.1$^{\circ}$ in size on the sky and compile a list of 70 clouds.  The physical size of each cloud is estimated based on the kinematic distance.  The distance to each cloud is found using the observed radial velocity, assuming a \citet{1993A&A...275...67B} rotation curve with the IAU-recommended Galactic center distance of 8.5 kpc and LSR orbital velocity of 220 km s$^{-1}$.  There are two kinematic distances for each negative radial velocity as discussed in $\S$ 4.1.  Since it is more likely that the cloud is at the near kinematic distance, we eliminate the far kinematic distance from the following calculations.  The results are listed in Table 2.  Column 1 gives the cloud number.  Columns 2 and 3 list the Galactic longitude and latitude respectively.  The radial velocity of each cloud is listed in Column 4 and the near kinematic distance, D$_{near}$ is given in Column 5.  The linear size of each cloud is found from the distance and size of the cloud on the sky and is listed in Column 6.     

The distance estimates are limited by cloud random motions.  The cloud-to-cloud velocity dispersion is approximately 6.9 km s$^{-1}$ \citep{1984A&A...136..368B}.  Distances for clouds with small radial velocities that are within twice the velocity dispersion (2$\sigma$ = 13.8 km s$^{-1}$) of 0 are listed as upper limits with the upper limit distance given by

\begin{equation}
D_{upper} = \frac{13.8 \ km \ s^{-1}}{A*\sin(2l)}
\end{equation}

\noindent where $A$ is Oort's first constant of 14.5 km s$^{-1}$ kpc$^{-1}$ and $l$ is the galactic longitude.  For most clouds, the uncertainty in the distance estimate is typically 400 - 700 pc.  Distances to clouds at longitudes nearing $l$ = 270$^{\circ}$ become very uncertain as the sin(2$l$) term in Equation 5 goes to zero.  For these clouds we place an upper limit distance of 10 kpc.  Clouds in the fourth quadrant of the Galaxy with small positive radial velocities, such as clouds \#44 and \#53 for example, probably only 'appear' to be on the far side of the solar circle due to cloud random motions.  Also, the Sun is surrounded by the Local Bubble, a region of very low neutral gas density which extends between 65-250 pc depending on the direction \citep{1999A&A...346..785S}.  Although the inner 1-2 pc of the Local Bubble is warm neutral medium \citep{2002ApJ...565..364S}, it is probable that any cold HI clouds are beyond 200-300 pc.  All upper limit distances are labeled by an asterisk in Table 2.  For these clouds the resulting size is also an upper limit.  

\subsection{Cloud Densities}

The column density, $N_H$, of the cool phase HI is determined by the
spin temperature, $T_s$, and the assumed optical depth, $\tau=1$, as

\begin{equation}
N_{H}  \ = \ C \ T_s \ \Delta v
\end{equation}

\noindent where $\Delta v$ is the equivalent width of the line, $\Delta v
= \int{\tau (v) dv}$, and $C$ = 1.823 $\times$ 10$^{18}$ cm$^{-2}$ K$^{-1}$ (km s$^{-1}$)$^{-1}$.
For the upper limit value of $T_s$ corresponding to 
$\tau \rightarrow \infty$ the column density also goes to 
infinity, so this is not useful for estimating the density.  
Values of $T_s$ are given on Table 2, Column 7 (for $\tau=1$)
and Column 8 (upper limit, $\tau \rightarrow \infty$), $\Delta v$
is given on Column 9, and Column 10 gives $N_H$ (assuming $\tau=1$
as in Equation 6).

We can translate the column density into a lower limit for the space
density, $n_{min}$, using the diameter, $d$, of the self-absorption detected
by the search routine of $\S$ 3 given the kinematic distance, $D$.
Assuming a uniform sphere gives $N = d \times n_{min}$.  This is a lower limit
for $n$, since the cloud geometry is far from spherical, and $d$ represents
the longest extent of the contiguous absorption on the plane of the
sky.  For $n_{min}$, a lower limit HI mass for each cloud can be determined.  Excluding clouds \#1 through \#15, whose distances and densities are highly uncertain, cloud HI masses range from 3 $M_{sun}$ (cloud \#27) to 4400 $M_{sun}$ (cloud \#33).   As can be seen on Figures 1, 2, 6, and 7, the HISA distribution
is very irregular on small scales, giving the impression of a very
low covering factor for the cool HI.  The line of sight filling
factor is probably also small, so that the effective path length
through the cool HI is much less than $d$, giving a higher $n$ for
a given observed $N$.  The lower limit density values $n_{min}$, are given
on Table 2, Column 11 and plotted on the upper panel of Figure 8
vs. $T_s$.  Lines on this figure show constant HI pressure,
$n T_s$, of 10 and 100 K cm$^{-3}$.  These pressure values are 
very low compared with other estimates of the kinetic pressure
in the interstellar medium which typically give $P$ $\ge$ 2000 K cm$^{-3}$ \citep{2001ApJS..137..297J}.

Estimates of the density in cool phase HI often give paradoxically
low pressures like those represented by the densities on Figure 8 (upper panel).  \citet{2003ApJ...586.1067H} find a similar result, which they interpret
as evidence for a thin sheet geometry for the absorbing gas, with
aspect ratios (sheet diameter to thickness) of 70 to a few hundred.
Our HISA clouds do not look like sheets, but rather like a collection
of clumps with low filling factor.  Geometrically this converts the
sheet into a sponge, with a large extent on the sky, $d$,  but
a small line of sight path length through the cool gas, $l \ll d$
so that the true density $n \gg n_{min}$.  Typically the filling
factor needed to bring the pressure above 2000 K cm$^{-3}$ is
$\sim 10^{-2}$, similar to the \citet{2003ApJ...586.1067H} estimate.
In the case of these HISA clouds, it is likely that there is some 
molecular hydrogen present as well, so the partial pressure of HI
may be only a small fraction of the total.  Thus filling factors
of 10$^{-1}$ or more are reasonable.  This matches the covering
factors indicated by the very spotty appearance of most of the
clouds.

\section{Discussions}
\subsection{Missing Link Clouds}
By determining the physical parameters (spin temperature, density, etc.) of HISA features throughout the SGPS, we can begin to establish how HISA fits in with other cold components of the ISM and develop a better understanding of the ISM as a whole.  From Table 2, the spin temperatures range from 6-41 K, assuming an optical depth of one.  For comparison, the upper limit spin temperature range for an infinite optical depth is 23-59 K.  This fits into the spin temperature gap between diffuse atomic clouds with spin temperatures of 25-100 K and dense molecular clouds with kinetic temperature, $T_k$ $\leq$ 10K, suggesting that gas in the interstellar medium (ISM) is not broken into discrete spin temperature groups, as our current observational tracers show, but instead has a continuous range of spin temperatures in the CNM from very cold to very warm.  Diffuse atomic clouds are defined to be those that show up in absorption towards continuum sources and dense molecular clouds are those that show CO in emission.

Table 2 gives the lower limit HI density for each cloud.  To determine the total density, $n_{HI}$ + $n_{H_2}$, we assume that each HISA cloud is in thermal pressure equilibrium with the surrounding ISM.  The average thermal pressure versus Galactic radius \citep{2003ApJ...587..278W} is given by

\begin{equation}
P_{th}/k \ = \ 1.4 \times 10^{4} \ exp(\frac{-R_g}{5.5}) \ K \ cm^{-3}
\end{equation}

\noindent where $P_{th}$ is the thermal pressure and $R_g$ is the distance from the Galactic center.  For the clouds listed in Table 2, the Galactocentric radius range used to get the pressure is 5.0 - 9.9 kpc.  The total hydrogen density, $n$$_{H}$ = $P$$_{th}$/$T$$_{s}$, ranges from 42 - 550 cm$^{-3}$, two orders of magnitude higher than the lower limit HI densities given in Table 2.  As an example, cloud \#32 ($R_g$ $=$ 7.0 kpc) has an HI pressure of 36 K cm$^{-3}$, 0.9\% of the expected pressure of 3900 K cm$^{-3}$.  Either the total density of the cloud is a factor of a hundred higher, in which case it is almost all molecular, or the HI is arranged in a sheet or a sponge with some filling factor that results in a higher HI density.  A plot of partial HI pressure versus observed $^{12}$CO brightness temperature for all 70 HISA clouds is shown in Figure 8.  Clouds with a higher CO brightness temperature tend to have low HI pressures.  This suggests that the total thermal pressure is, at least in some part, supported by molecular gas.  Considering the HI component only, the ISM pressure would have to be extremely low for these clouds to be homogeneous and in dynamical equilibrium with the outside.  Alternatively, these clouds may not be in pressure equilibrium with the global ISM.  The HI pressure of our HISA couds is well below what is expected of the surrounding ISM, but it is likely the pressure is higher than the reported HI pressures.  This is supported by the presence of CO emission in many clouds.  There is also likely to be emission from other molecules.  Those HISA clouds that lack corresponding CO emission, may show a correspondence with other molecular tracers.  Observations of other molecules, such as OH and CH, are needed to further address this issue.  

With the SGPS, we can study this spin temperature and density regime of the ISM.  HISA clouds provide an observational tool to study the least molecular of the molecular clouds.  The resulting temperature and density ranges for the HISA clouds studied in this paper show that HISA represents a population of missing link clouds that fill in the temperature and density gaps between diffuse atomic gas and dense molecular gas.  The relationship between atomic and molecular gas in these clouds is discussed further in the next section.

\subsection{Comparison to $^{12}$CO}

We compare our HI emission maps with $^{12}$CO emission maps from \citet{1987ApJ...322..706D} and \citet{2001ApJ...547..792D} and find that 60\% of all detected HISA features have a $^{12}$CO brightness temperature of at least 0.5 K.  Figure 6 shows two examples.  Cloud \#42 from Table 2 shows a very clear connection between the atomic gas and the molecular gas.  This cloud, located at $l$=320.1$^{\circ}$, $b$=$-$0.4$^{\circ}$, and $v$=$-$8.24 km s$^{-1}$ is completely enveloped with $^{12}$CO contours, having a peak $^{12}$CO brightness temperature of 1.75 K.  However, for cloud \#60 the connection is not as clear.  This cloud, located at $l$=336.6$^{\circ}$, $b$=$-$0.8$^{\circ}$, and $v$=$-$26.37 km s$^{-1}$ shows only a small amount of diffuse $^{12}$CO emission peaking at 0.3 K.  Although this cloud is not at the center of a concentration of molecular gas, the poor resolution of the $^{12}$CO data does not eliminate it from being associated.  The HI could be outside of the molecular cloud, but connected as part of a surrounding HI shell.  Figure 7 shows an example of a HISA cloud, Cloud \#10 from Table 2, that has no associated molecular emission above the rms noise level of the survey.   

The comparison between HISA and $^{12}$CO is illustrated on a broader scale in Figures 9 and 10.  The top panel in each Figure is an $l-v$ diagram showing the location of all detected HISA features in l-v space for a 5$^{\circ}$ strip in longitude in the Galactic plane, with $^{12}$CO contours.  The bottom panel is the corresponding $^{12}$CO emission.  In both images there are many cases where there is a clear correlation between HISA and molecular emission, and cases where there is no correlation.  For example, in Figure 9, HISA detected near $l$=327.3$^{\circ}$ and $v$=$-$10 km s$^{-1}$ shows no corrleation while gas at the same longitude and $v$=$-$45 km s$^{-1}$ shows a very clear correlation.  There are many other examples of both cases in Figures 9 and 10 and throughout the SGPS.  

Why is it that some HISA features are associated with molecular gas while others are not?  It is possible and has previously been proposed that HISA clouds could be in a transtion from an atomic state to a molecular state, or vice versa \citep{2001ApJ...555..868M, 2002ASPC..276..235G, 2003ApJ...598.1048K}.  \citet{1992A&A...253..525A} and \citet{1993ApJ...411..170E} discuss the H/H$_{2}$ transition in interstellar clouds and galaxies and show that the transition time can be quick (10$^{6}$ yr) and requires only slight changes in the radiation field or ambient pressure.  If a molecular cloud, possibly with traces of atomic gas due to cosmic ray ionization, has recently undergone star formation or enters a region of newly formed stars, it will experience an increase in the radiation field.  The resulting photodissociating UV radiation can penetrate the cloud and destroy H$_{2}$ molecules.  If the destruction rate exceeds the rate at which the HI collects on grains to form H$_{2}$, the cloud will make a transition from molecular to atomic.  From the models of \citet{1992A&A...253..525A}, for a cloud of density $n$=100 cm$^{-3}$, typical of the total number densities of our clouds if they are in pressure equilibrium with the ISM, the HI and H$_{2}$ abundances are equal at a cloud optical depth of 0.03.  At this cloud depth the photodissociation rate is 5 $\times$ 10$^{-15}$ s$^{-1}$.  The atomic abundance in the cloud begins to dominate when the photodissocation rate reaches 10$^{-11}$ s$^{-1}$.  This would likely be the case if all of the clouds listed in Table 2 had densities a factor of 100 higher.  This corresponds to a sheet-like cloud structure.  On the other hand, if the clouds are all uniformly spherical in structure, they will have the lower limit densities given in Table 2 and be dominated by molecular gas.  Such clouds would have a much lower photodissociation rate of $\le$ 10$^{-17}$ s$^{-1}$.   If the cloud is mostly atomic and becomes shielded by dust, the reverse effect can occur.  UV radiation will not penetrate the cloud and H$_{2}$ molecules will form, resulting in a transition from atomic to molecular.  Another way to initiate a molecular/atomic transition is through the interaction with spiral arms. This is discussed further in the next section.  Based on the CO emission maps, it is probable that our HISA clouds contain a wide range of molecular gas fractions, such that some have higher atomic abundances while others have higher molecular abundances and we are thus seeing the clouds at different transition stages.  

\subsection{Spiral Arms}

Is there a change in the HISA-CO correlation between spiral arm and interarm regions?  If we can detect a correlation or anti-correlation between HISA and CO, given the cloud location relative to the spiral arms, we can determine the process that initiates a phase transition and direction of the transition for a specific cloud.  \citet{1987ApJ...317..646P} show that in their observations, $^{12}$CO and HISA are well correlated within the spiral arms, but in the inter-arm region there is actually an anti-correlation.  This can be explained if HI clouds interact with a spiral density wave within a spiral arm.  As the atomic gas crosses the arm, the increase in pressure converts HI into H$_{2}$.  The timescale for this transition can be very short, on the order of a few million years \citep{1993ApJ...411..170E, 2002PASJ...54..223H}.  Thus the correlation of cold atomic and molecular gas in spiral arms is not unexpected.  As the newly formed molecular cloud leaves the density wave, molecular gas will convert back into atomic gas as star formation is triggered.  In this model, HISA clouds in the interarm regions are more likely to be found without a molecular counterpart as the molecular gas is potentially all converted to atomic.  HISA tracing spiral arm structure has previously been proposed for individual clouds \citep{1976A&A....49..343S, 1992AJ....103.1627S} and for multiple cloud distributions \citep{2004Gibson}. 

Is this seen in the SGPS?  An $l-v$ diagram ranging from $l$=298$^{\circ}$ to $l$=318$^{\circ}$ is shown in Figure 11.  Galactic spiral arms \citep{2002astro.ph..7156C} are plotted on the top panel as well as $^{12}$CO emission on the bottom.   This model describes the electron density distribution of the Galaxy based on the spiral model of \citet{1976A&A....49...57G} from HII regions and radio continuum features and is the most detailed model of spiral arm position available.  The spiral arms plotted in Figures 11 and 12 are a transformation of the ($R$, $\theta$) loci in the \citet{2002astro.ph..7156C} model to ($l$, $b$) space, assuming a simple rotation law.  Both the far-side and near-side of each arm is shown in our figures for completeness.  However, only the near-side of the arms (v $\sim$ $-$50 to $-$60 km s$^{-1}$ for the Scutum-Crux arm and v $\sim$ $-$20 km s$^{-1}$ for the Sagittarius-Carina arm) can accurately be compared to HISA, since HISA is much more likely to be detected at close distances.  Velocity perturbation of the spiral arms also effects the appearance of the $l-v$ diagrams in Figures 11 and 12.  Preturbations can alter the location of the spiral arms and departures from circular rotation can effect the relative cloud positions by as much as $\sim$ 10 km s$^{-1}$.  This, along with other concerns discussed in this section, places limits on our ability to match HISA clouds with spiral arm and inter-arm regions.      

At the longitude where the Scutum-Crux arm is tangent to the line of sight, $l$=310$^{\circ}$ to $l$=312$^{\circ}$, there are a large number of HISA pixels, all of which are associated with molecular gas.  This is consistent with the model discussed above.  However, there are regions on the $l-v$ diagram that are not consistent with this model.  There is some HISA near the Sagittarius-Carina arm at a velocity of approximately 20 km s$^{-1}$ but little of it is correlated with molecular gas.  This is not necessarily unexpected.  At this velocity, the arm is at a galactocentric radius of approximately 9 to 10 kpc where the molecular gas emission is expected to drop off.  The cooling mechanism for HISA clouds absent of CO emission at large radii and in interarm regions is unkown, but the cold HI may still be associated with molecules that are unobservable by current surveys.  Although it is possible that the associated groups of atomic and molecular gas at longitudes of $l$=298$^{\circ}$ to $l$=312$^{\circ}$ at a velocity of $-$20 to $-$45 km s$^{-1}$ are connected with the Sagitarius-Carina arm, the connection is not entirely clear.

Velocity perturbation of the spiral arms can play a role in this.  As atomic clouds enter a spiral arm and make a transition to molecular gas, a warm background of HI emission piles up behind the cool clouds at the same radial velocity, thus making the HISA phenomenon observable.  This has been observed in the Perseus arm \citep{2004Gibson}. Spiral density waves are required to explain the presence of strong HISA features within the arms.  In the case of the Perseus arm, HISA is seen downstream of the Perseus arm shock, where HI is cooling rapidly and is potentially on the way to forming new stars \citep{2002ASPC..276..235G}.  The HISA features in Figure 11 near the Sagittarius-Carina arm around $-$40 km s$^{-1}$ could be clouds that have left the arm and are making the transition back to atomic gas.  A second $l-v$ diagram is shown in Figure 12 with similar conclusions.  At the longitude where the spiral arm is tangent to the line of sight, at $l$=284$^{\circ}$, there is a grouping of atomic and molecular gas that appears to be within the arm.  But at other locations along the arm, there appears to be very little connection.  This is further shown in Figure 13.  Here, the Galactic spiral arms \citep{2002astro.ph..7156C} are plotted on the Galactic plane as viewed from above, with the coordinate origin corresponding to the Galactic Center, as well as the HISA clouds listed in Table 2, excluding those within 10 degrees of $l$=270$^{\circ}$ due to large uncertainties in the distance.  HISA clouds that show a clear connection with $^{12}$CO are plotted as filled squares and those with no molecular counterpart are plotted as open squares.  Most of the clouds are located near the Sun since it was assumed earlier that HISA is more likely to lie at the near-kinematic distance.  

There are a set of HISA clouds corresponding with CO emission near the Scutum-Crux arm in Figure 13.  This is expected from the model described before, but there are also a set of clouds corresponding with CO emission that are clearly in the inter-arm region between the Scutum Crux and Saggitarius-Carina arms ($x$ $=$ $-$2.7 kpc, $y$ $=$ 6.5 kpc).  There are also clouds, showing no correspondence with CO, near the Saggitarius Carina arm ($x$ $=$ $-$0.75 kpc), contradicting the described model.  The results presented in Figure 13 suggest that HISA correlated with CO is weakly constrained to the spiral arms, but a correlation is also present in the inter-arm regions.  There are a few possible explanations for this.  It could be that the coarse angular and velocity resolution of the $^{12}$CO data prevents us from accurately matching with the HI data.  It may also be that matching $^{12}$CO with HI data is not the best approach to studying the atomic/molecular transition in such clouds.  At these densities the excitation of the J = 1-0 transition of CO may be subthermal, and CO emission is probably not accurately tracing the molecular gas.  Also, the plotted spiral arms are based on a model that is poorly constrained and still uncertain.  Regardless, there is no conclusive evidence that HISA with a molecular counterpart is constrained to the spiral arms.

\section{Conclusions}

We have presented an automated search routine to look for neutral hydrogen self absorption clouds in the Southern Galactic Plane Survey.  The routine takes advantage of the characteristic absorption dip seen in HISA.  By looking at the first and second derivatives of the velocity profiles with respect to the brightness temperature, self-absorption clouds can easily be found, even those that would normally be undetectable by the eye alone.  We search the entire SGPS for self-absorption features and derive spin temperature ranges for each SGPS subset.  We select 70 of the larger features finding spin temperatures that range from 6-41 K with total number densities of a few hundered cm$^{-3}$.  These missing link clouds fill in the spin temperature and density gap between diffuse atomic clouds and dense molecular clouds, providing an observational tool to study this regime of the ISM.  Instead of having specific temperature and density ranges, the cold gas in the ISM has a continuous distribution.  Comparing HISA with $^{12}$CO emission shows 60\% of detected HISA is correlated in space and in velocity with molecular gas.  We propose that this correlation is due to an atomic/molecular phase transition.  The clouds analyzed show a wide range of atomic and molecular abundances which is dependent on the nearby photodissociation rate.  Longitude-velocity diagrams show the distribution of HISA with a molecular counterpart versus that without a counterpart.  Although a correlation with the galactic spiral arms may be expected, no conclusive correlation is seen in the SGPS.  The coarse angular resolution of the $^{12}$CO data limits the comparison with HI.  Better molecular observations are needed, including those of other molecular tracers, such as OH and CH.  

We thank an anonymous referee, whose comments and suggestions improved the paper.  The Australia Telescope is funded by the Commonwealth of Australia for operation as a National Facility managed by CSIRO.  This research has made use of the Astrophysics Data System Abstract Service of NASA.  The CO data was provided by Dame et al. via the CDS Service for Astronomical Catalogues.  We thank Evan Skillman for helpful comments and discussions.  This work was supported by NSF grant AST-0307603 to the University of Minnesota.

\begin{figure}
\epsscale{1.0}
\plottwo{f1a.eps}{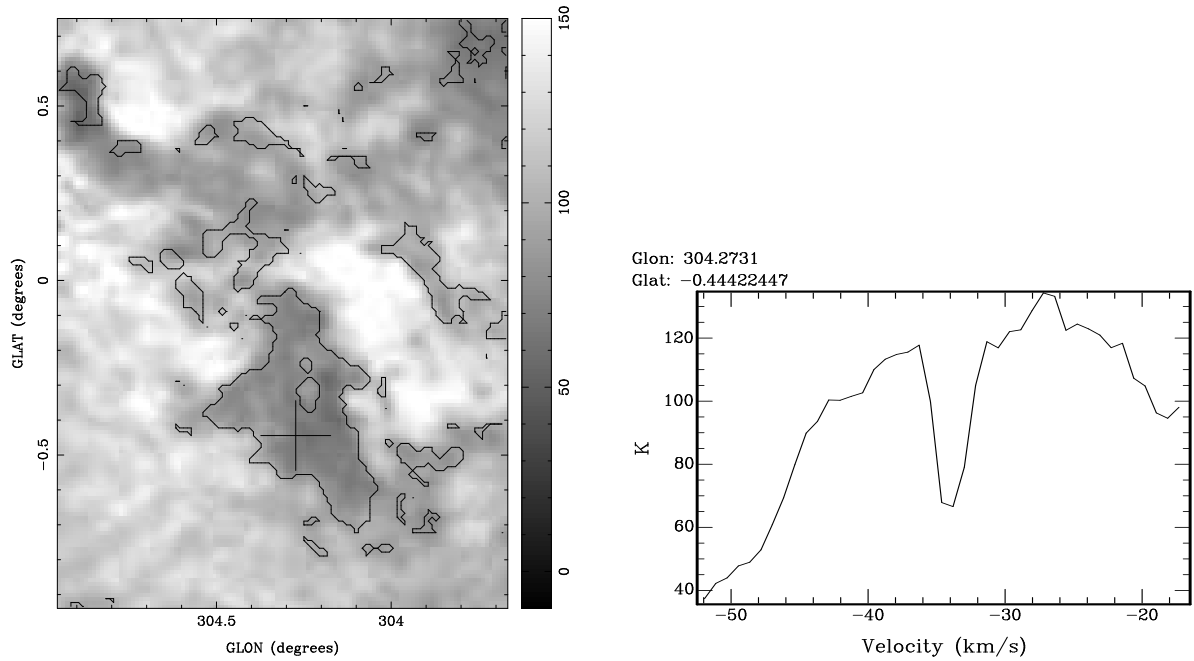}
\caption{Top:  A brightness temperature map (K) illustrating a complex of HISA clouds centered around $l$ = 304.2$^{\circ}$, $b$ = $-$0.1$^{\circ}$, and $v$ = $-$33.80 km s$^{-1}$.  The black contours indicate regions that have passed the HISA threshold test described in $\oint$ 3.    Bottom:  A velocity profile from one of the clouds in the complex at $l$ = 304.3$^{\circ}$ and $b$ = $-$0.4$^{\circ}$.  The cross in the top image marks this position.}
\end{figure}

\begin{figure}
\epsscale{1.0}
\plotone{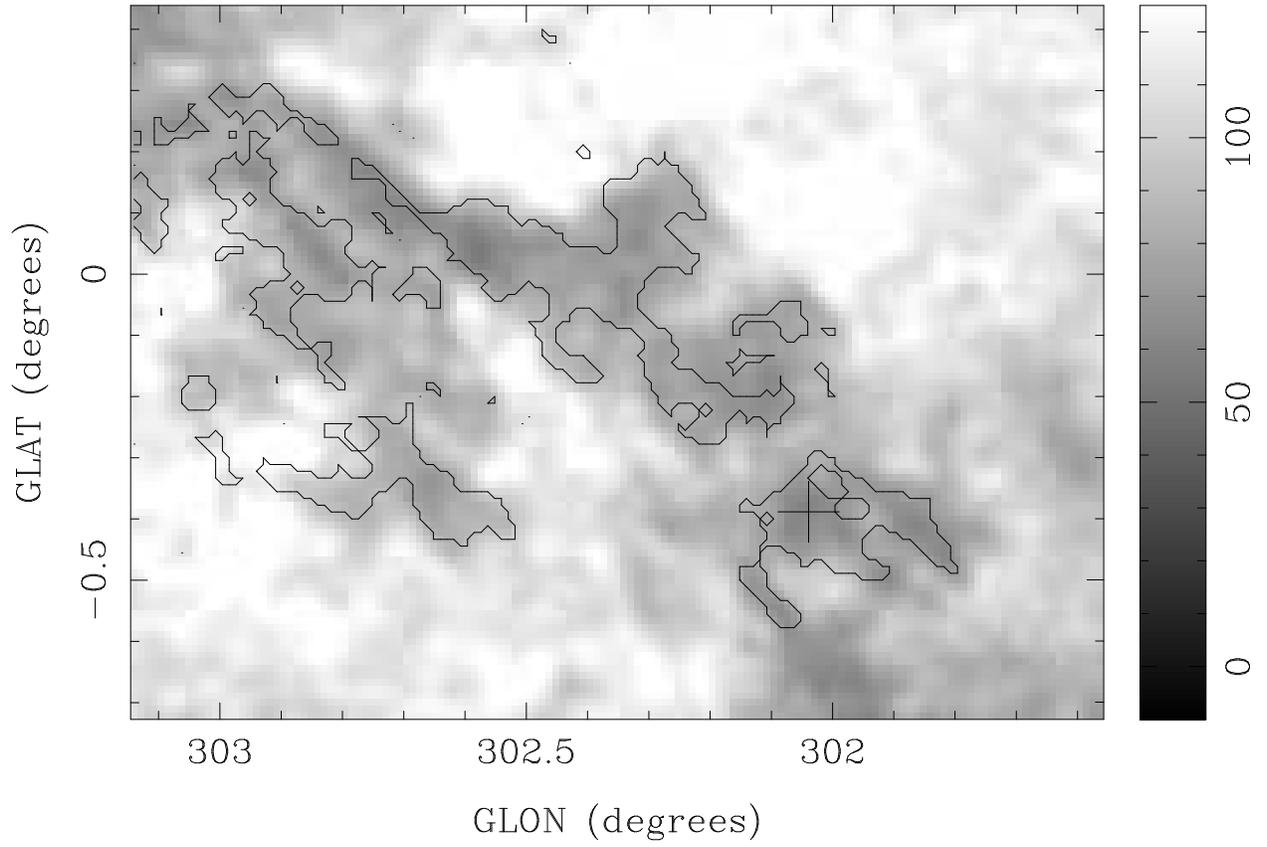}
\caption{A second brightness temperature map (K) showing a typical HISA cloud seen in the SGPS at a velocity of v = $-$34.63 km s$^{-1}$.   The cloud is seen as the dark region that extends from the upper left to the lower right of the image and is outlined by black contours.}
\end{figure}

\begin{figure}
\epsscale{0.5}
\plotone{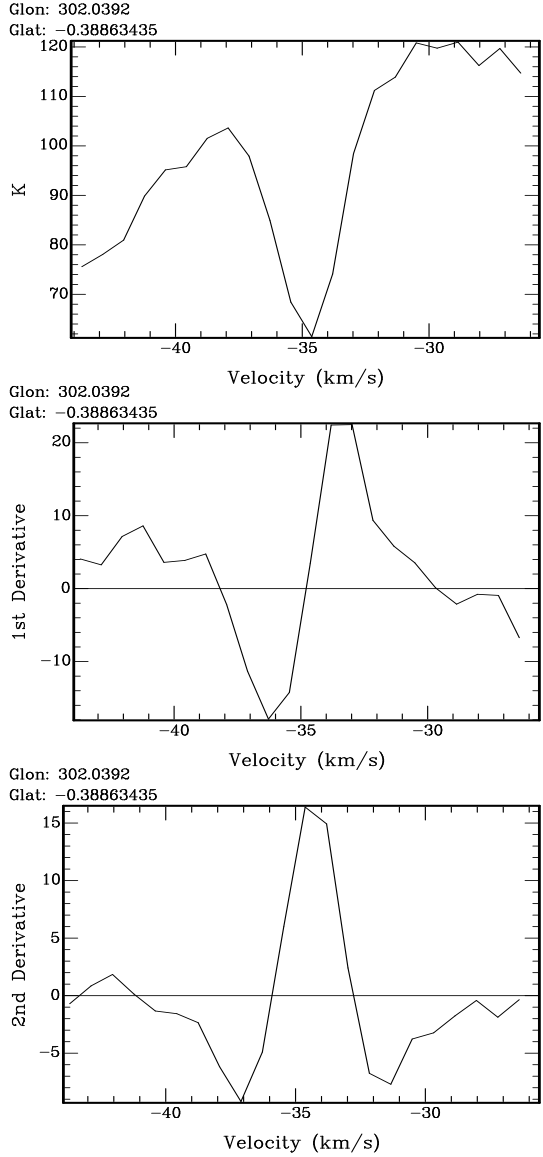}
\caption{Top:  A velocity profile of the cloud in Figure 2 at $l$ = 302.0$^{\circ}$, $b$ = $-$0.4$^{\circ}$, and $v$ = $-$34.63 km s$^{-1}$, marked by a cross.  Middle:  The first derivative of the brightness temperature profile in the top panel with respect to the radial velocity in units of K/(km s$^{-1}$).  The profile has a negative peak where the HI temperature begins to drop rapidly, crosses zero when the temperature reaches a minimum, and has a positive peak when the temperature starts to rise again.  Bottom:  The second derivative of the velocity profile in the top panel in units of K/(km s$^{-1}$)$^{2}$.  The second derivative profile contains a single positive peak representing the change between the negative peak and the positive peak in the first derivative profile.}
\end{figure}

\begin{figure}
\epsscale{1.5}
\plottwo{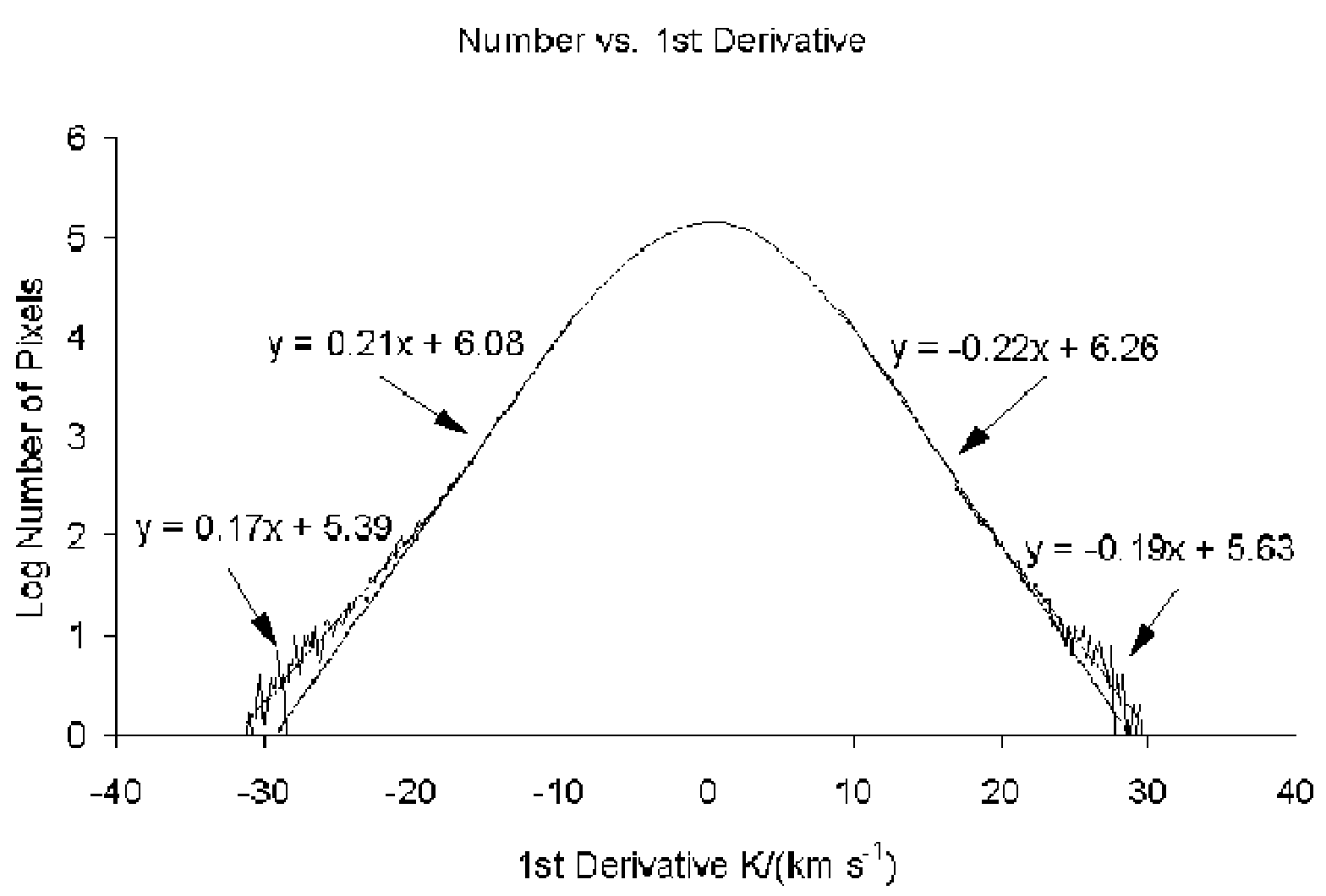}{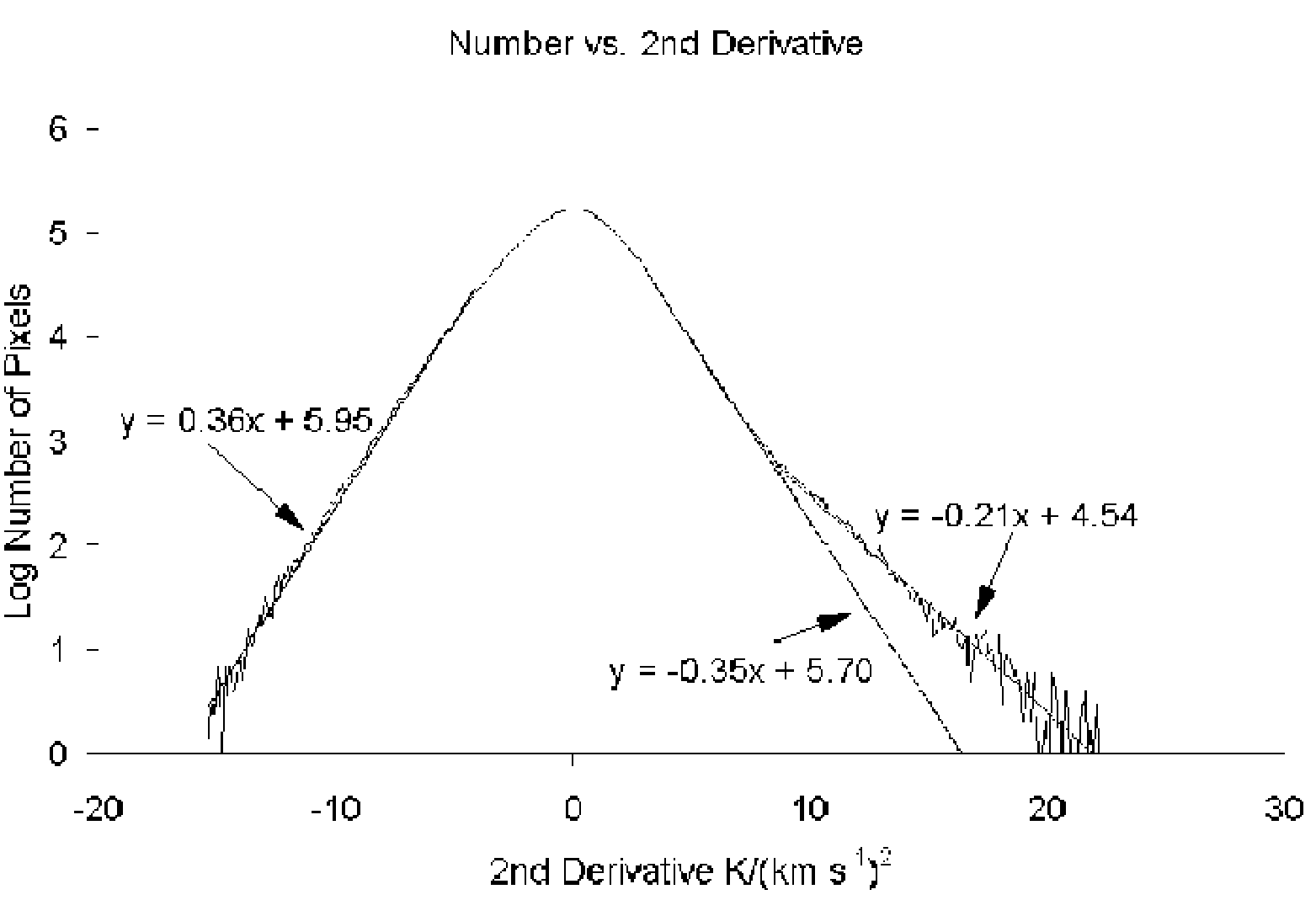}
\caption{Top:  The number distribution function of the values of the first derivative for a 5$^{\circ}$ strip of the Galactic Plane centered around $l$=330$^{\circ}$.  In log space, two linear lines can be fit on each side of the distribution.  The two fits with the smallest slope represent a population of HISA, due to the negative and positive peaks shown in Figure 3.  Bottom:  The number distribution function of the values of the second derivative.  In this distribution only the positive side can be fit with two lines.  The line with the lesser slope represents a population of HISA due to the single, positive peak of the second derivative profile in Figure 3.}
\end{figure}

\begin{figure}
\epsscale{0.7}
\plotone{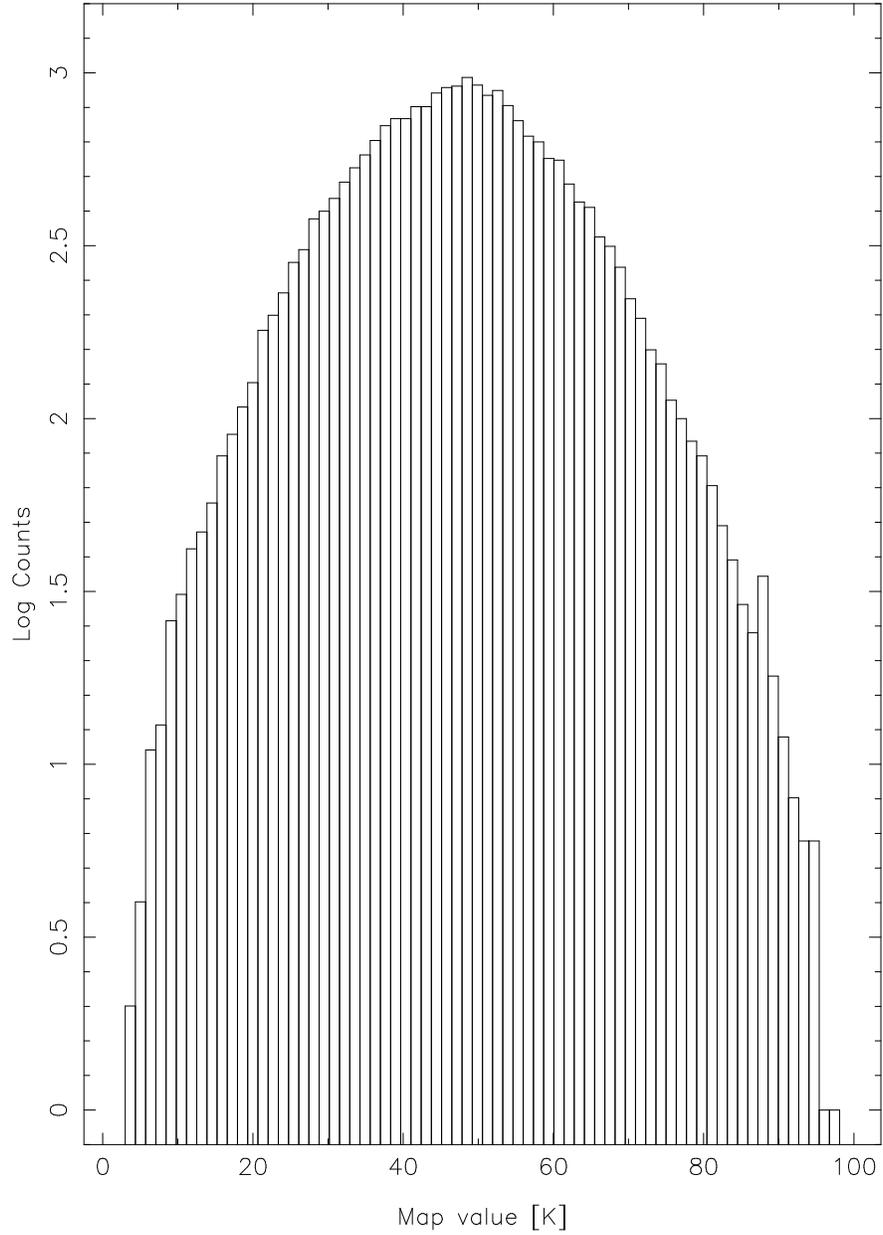}
\caption{A histogram of HISA pixels versus spin temperature for the SGPS data cube centered on $l$=310$^{\circ}$, showing a broad range of spin temperatures ($\tau$ = 1), including many that are very cool ($T_s$ $\le$ 20 K).}
\end{figure}

\begin{figure}
\epsscale{1.2}
\plottwo{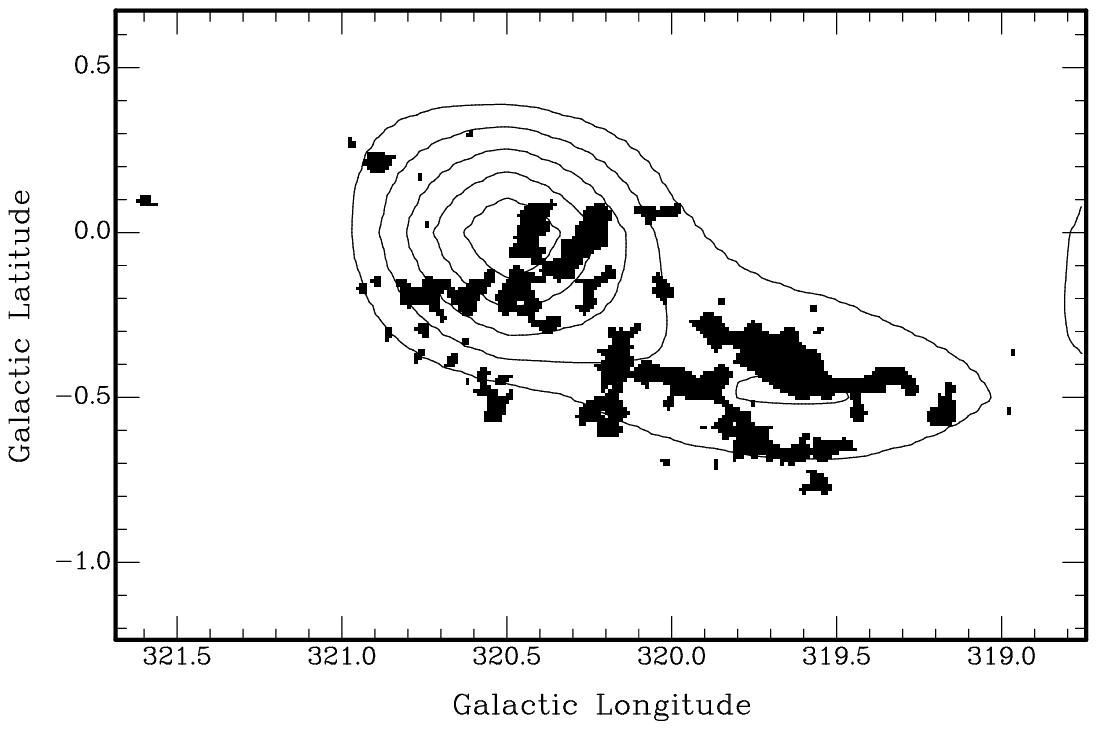}{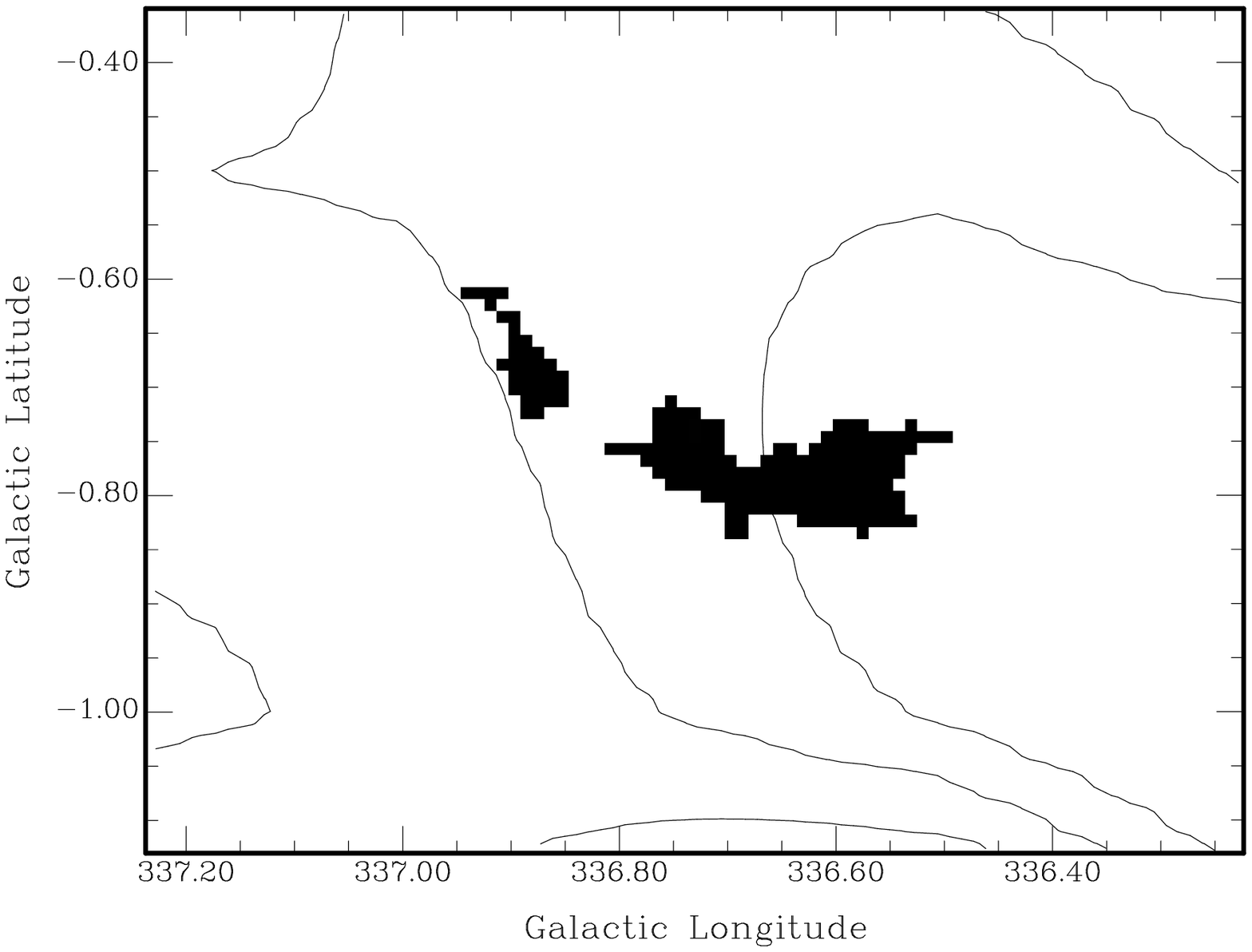}
\caption{Top:  A HISA cloud located at $l$=320.1$^{\circ}$, $b$=$-$0.4$^{\circ}$, and $v$=$-$8.24 km s$^{-1}$.  $^{12}$CO contours are labeled and range from 0.5K to 1.75K in increments of 0.25K.  The brightness temperature of the molecular gas peaks at 1.75K.  Bottom:  A second HISA cloud with CO contours ranging from 0.15 to 0.35K in increments of 0.1K.  This cloud is located at $l$=336.6$^{\circ}$, $b$=$-$0.8$^{\circ}$, and $v$=$-$26.38 km s$^{-1}$, and shows a minimal concentration of molecular gas, possibly associated with the outer envelope of a molecular cloud.}
\end{figure}

\begin{figure}
\epsscale{0.6}
\plotone{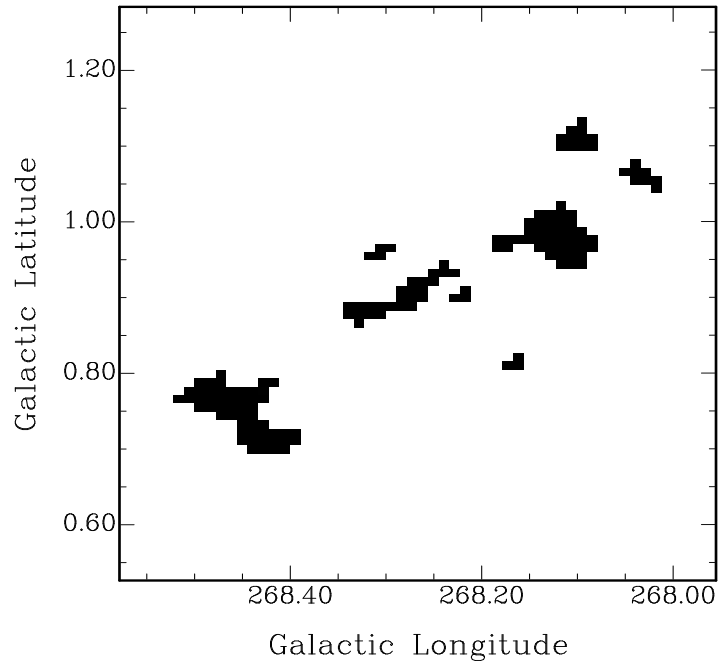}
\caption{A HISA feature located at $l$=268.2$^{\circ}$, $b$=0.9$^{\circ}$, and $v$=$+$14.85 km s$^{-1}$.  In contrast to the HISA clouds in Figure 6, this cloud has no associated $^{12}$CO emission above the 0.12 K noise level of the survey. }
\end{figure}

\begin{figure}
\epsscale{1.5}
\plottwo{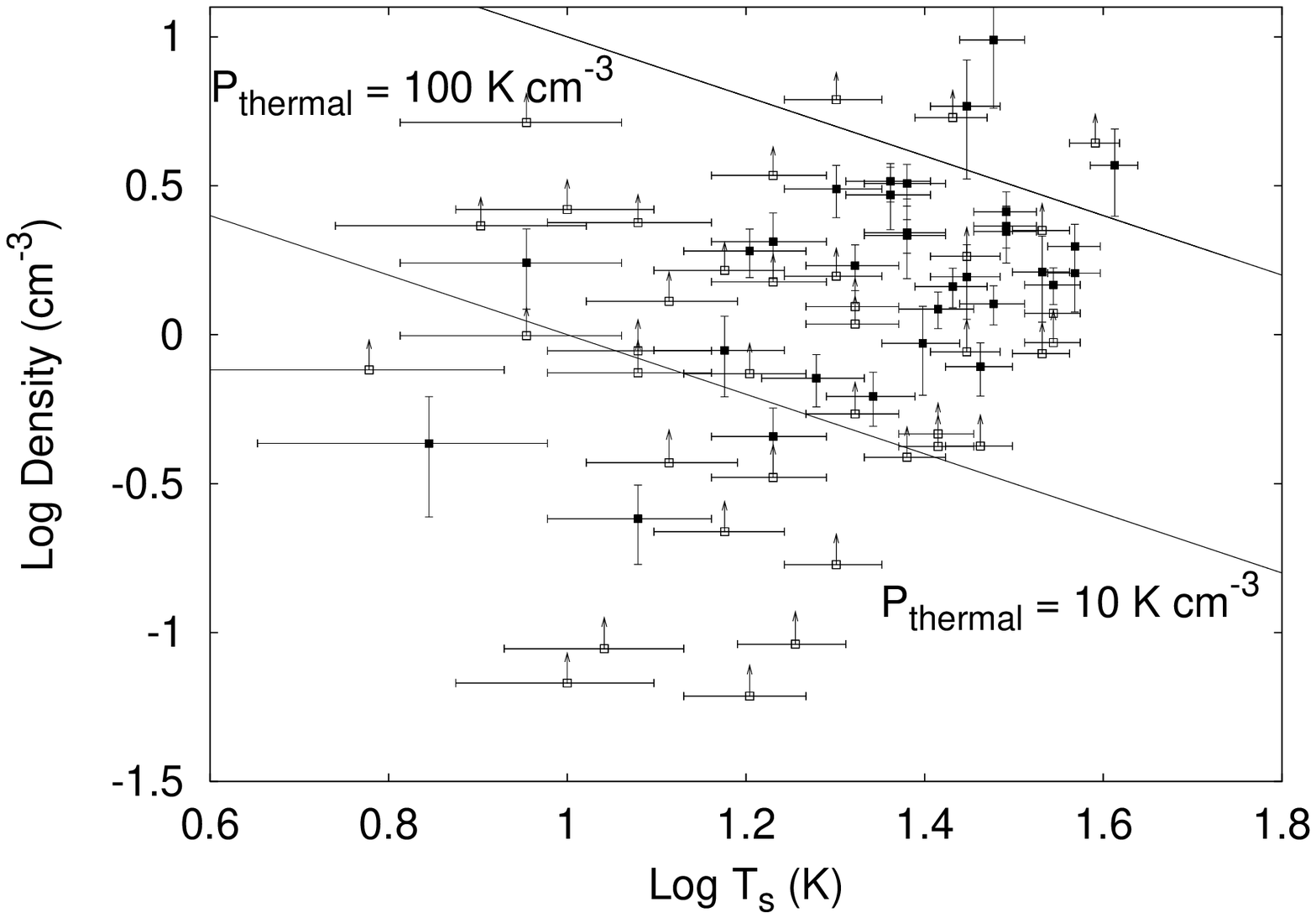}{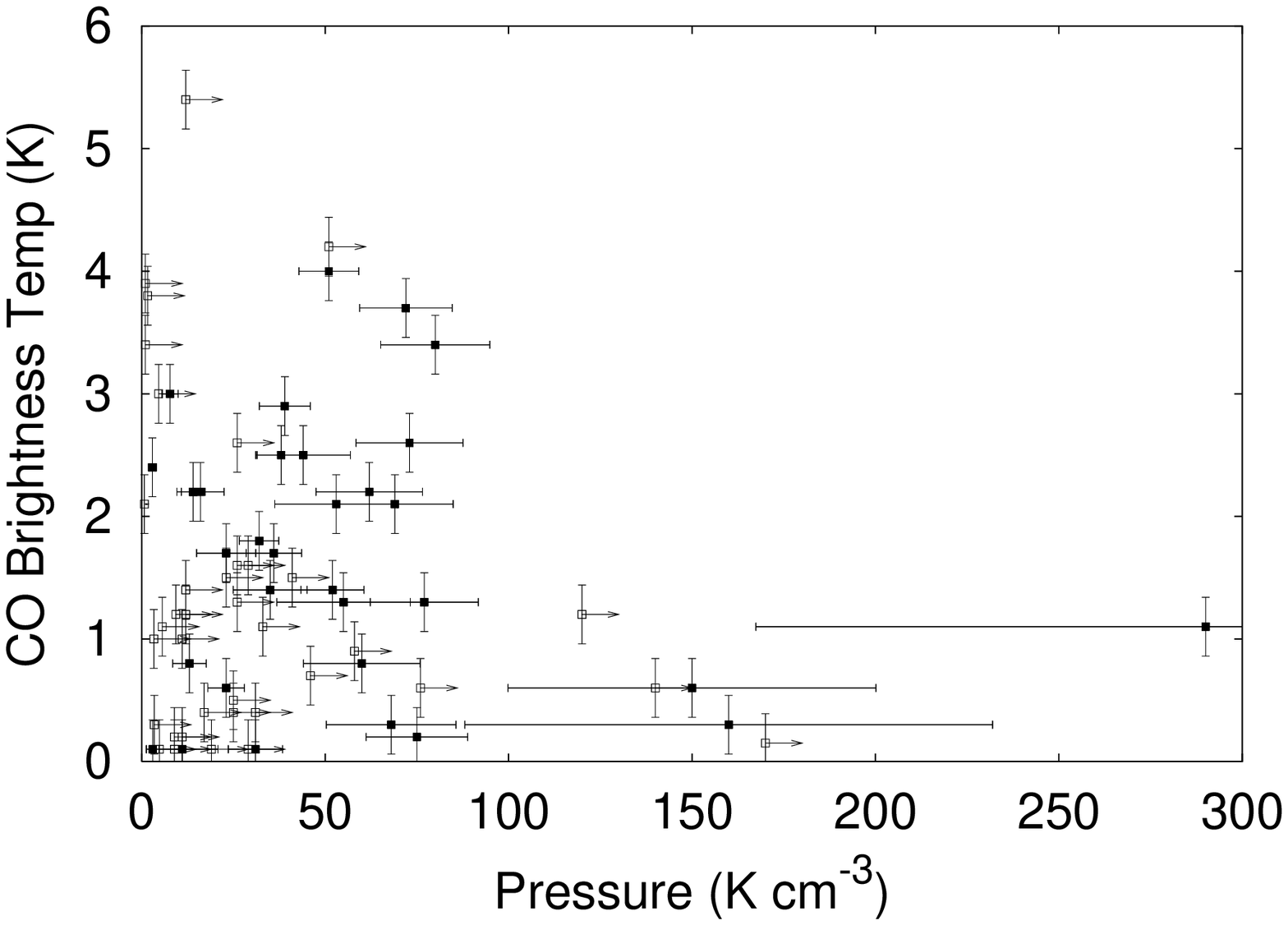}
\caption{Top:  A log-log plot of spin temperature versus HI number density for all 70 HISA clouds in Table 2.  Two lines of constant pressure, 10 K cm$^{-3}$ and 100 K cm$^{-3}$ are shown.  Open squares with arrows correspond to clouds whose distances are given as upper limits in Table 2.  The density estimate for these clouds is a lower limit.  Bottom: A plot of $^{12}$CO brightness temperature versus HI pressure.  Pressure estimates for the open squares with arrows are lower limits.}
\end{figure}

\begin{figure}
\epsscale{1.7}
\plottwo{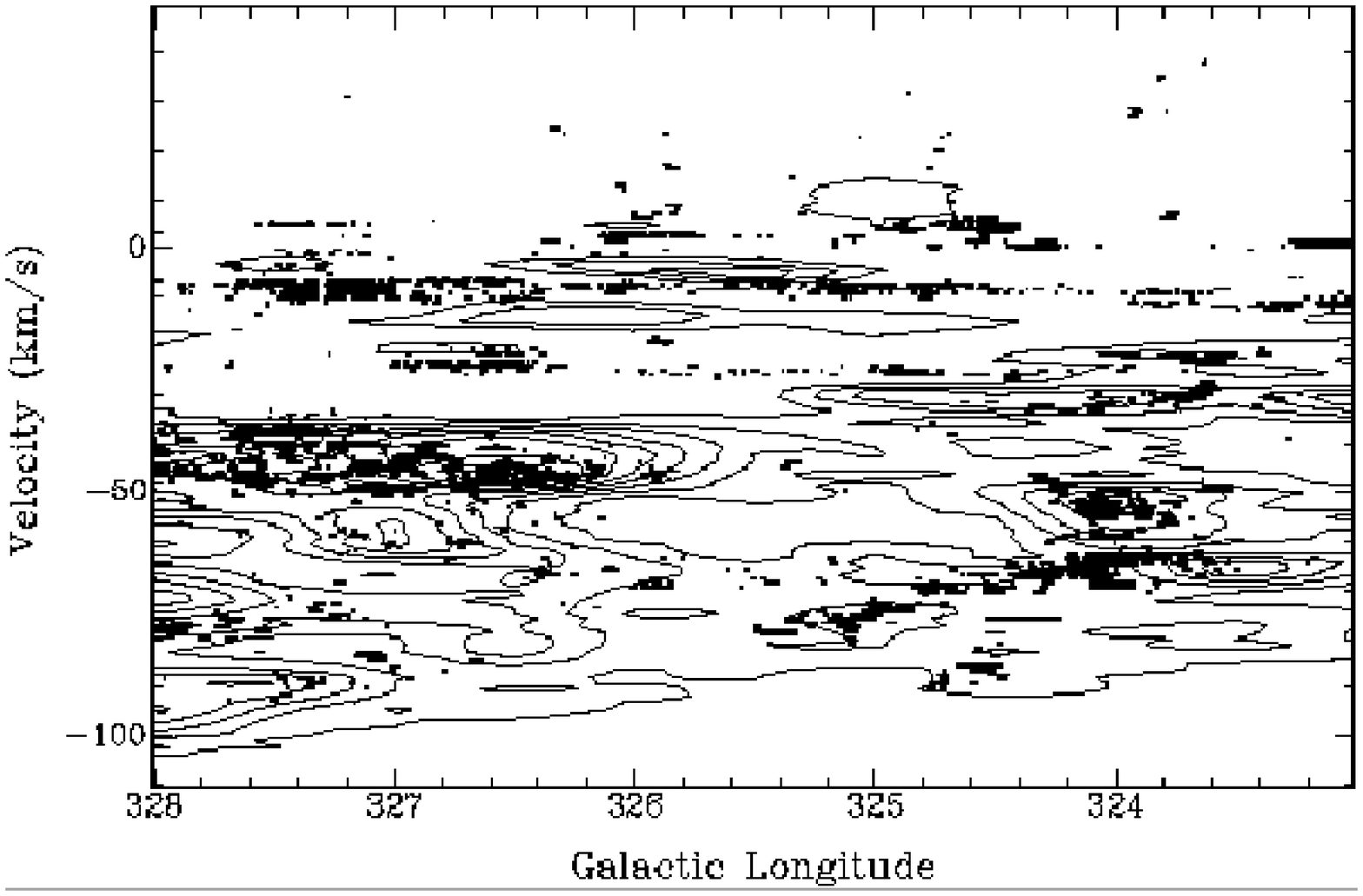}{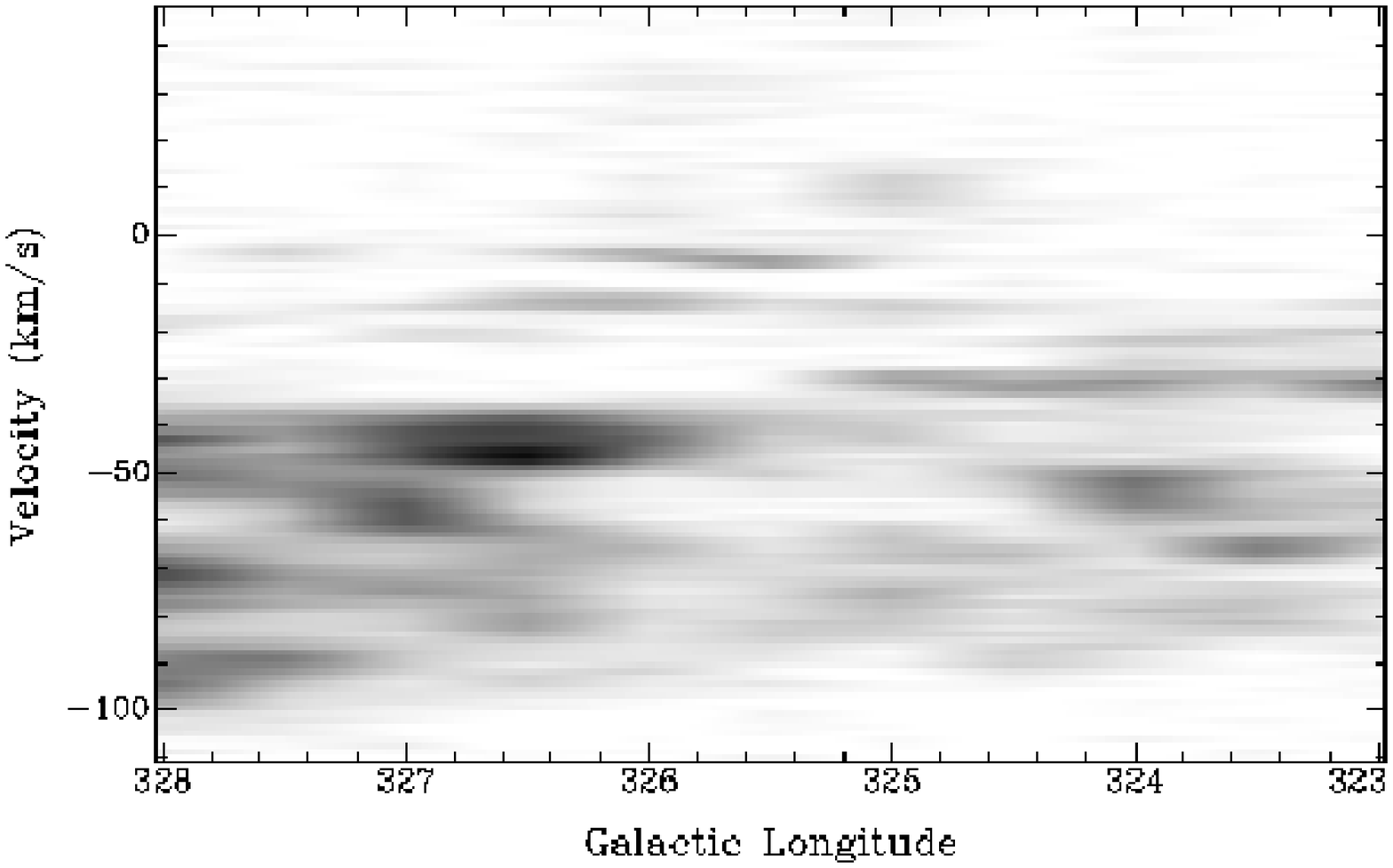}
\caption{Top: An $l-v$ diagram of the region ranging from $l$=323$^{\circ}$ to $l$=328$^{\circ}$.  $^{12}$CO contours are outlined in black and show that in some cases, HISA is associated with molecular gas, while in a number of other cases it is not.  The contour levels range from 0.5K to 4.5K in increments of 0.5K.  Bottom: A greyscale map of the $^{12}$CO emission with a temperature range of 0K (white) to 5K (black).}
\end{figure}

\begin{figure}
\epsscale{1.7}
\plottwo{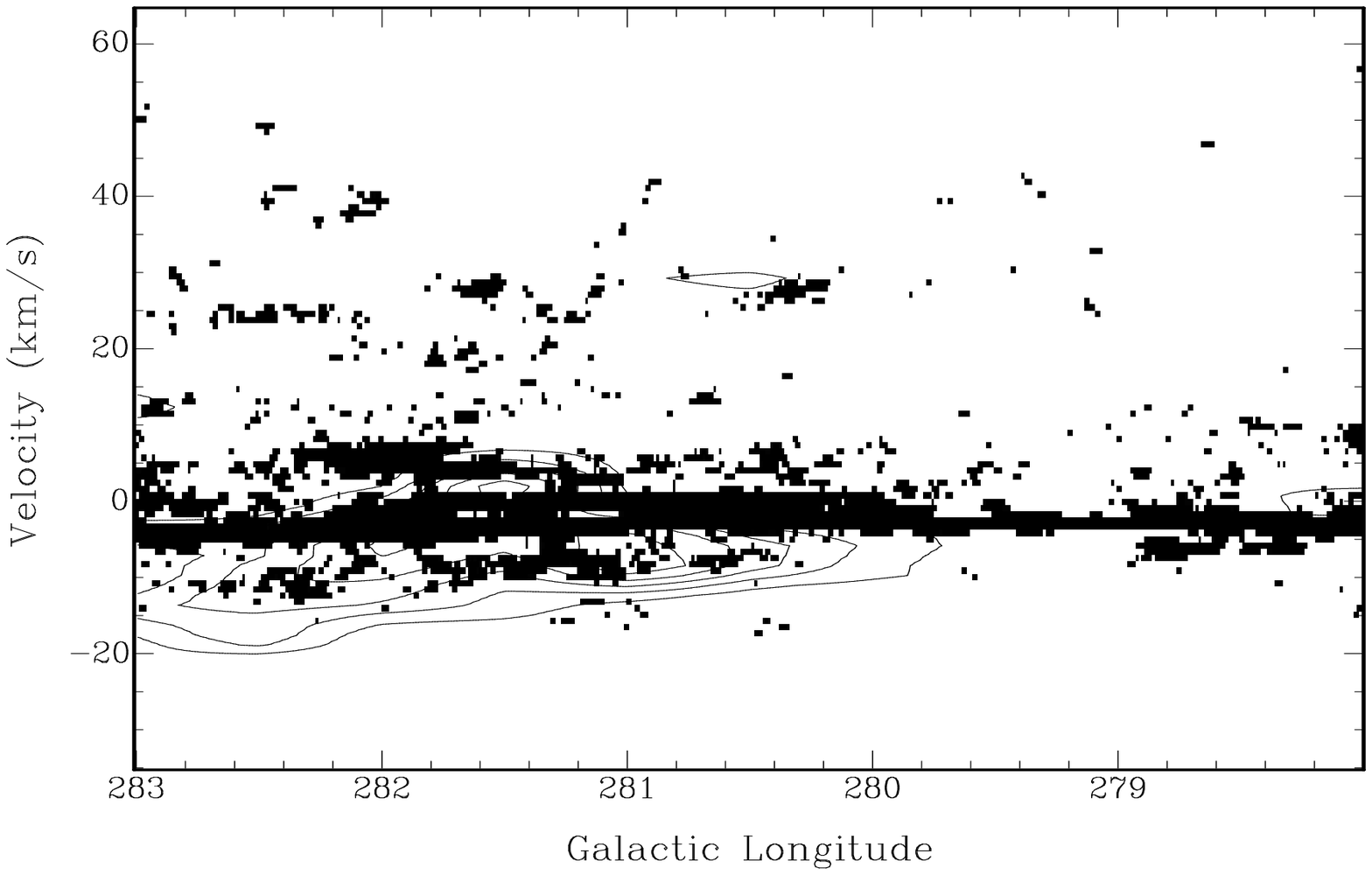}{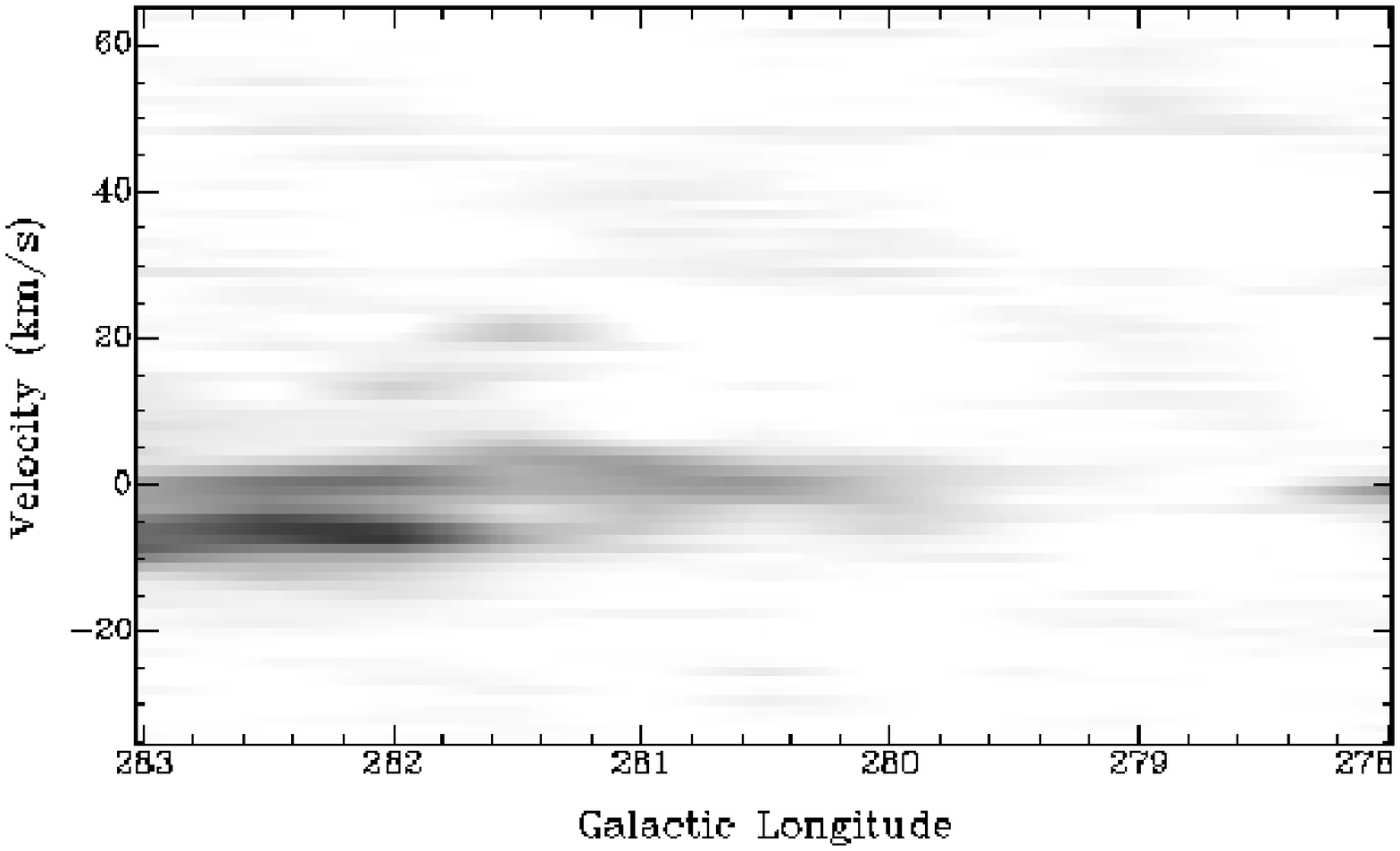}
\caption{Top:  An $l-v$ diagram of the region ranging from $l$=278$^{\circ}$ to $l$=283$^{\circ}$.  $^{12}$CO contours, ranging from 0.5K to 2.0K in increments of 0.5K, are outlined in black and just as in Figure 9, show that HISA is associated with molecular gas in some cases, but is anti-correlated in others.  Bottom:  A greyscale map of the $^{12}$CO emission with a temperature range of 0K (white) to 2K (black).}  
\end{figure}

\begin{figure}
\epsscale{1.7}
\plottwo{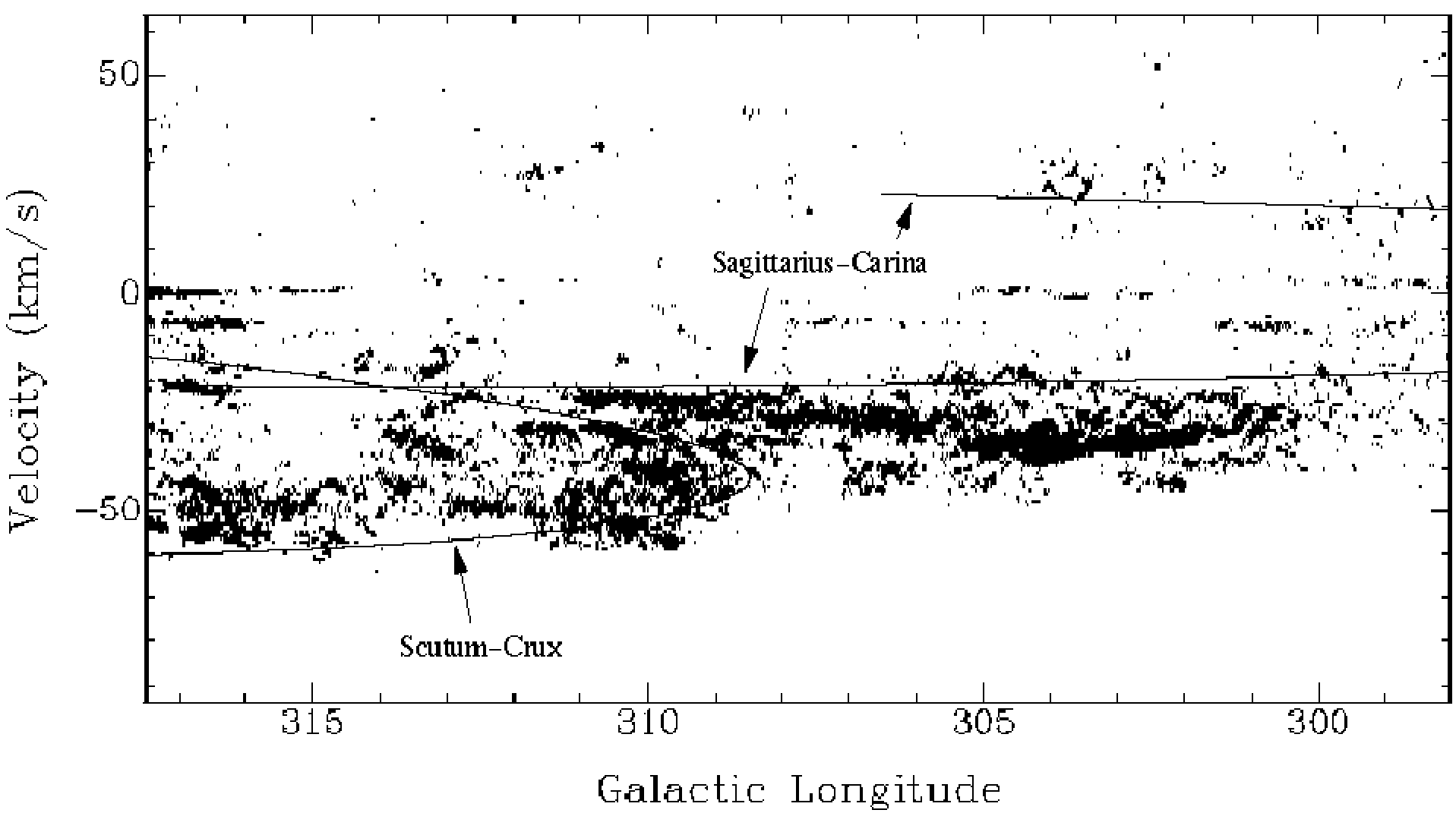}{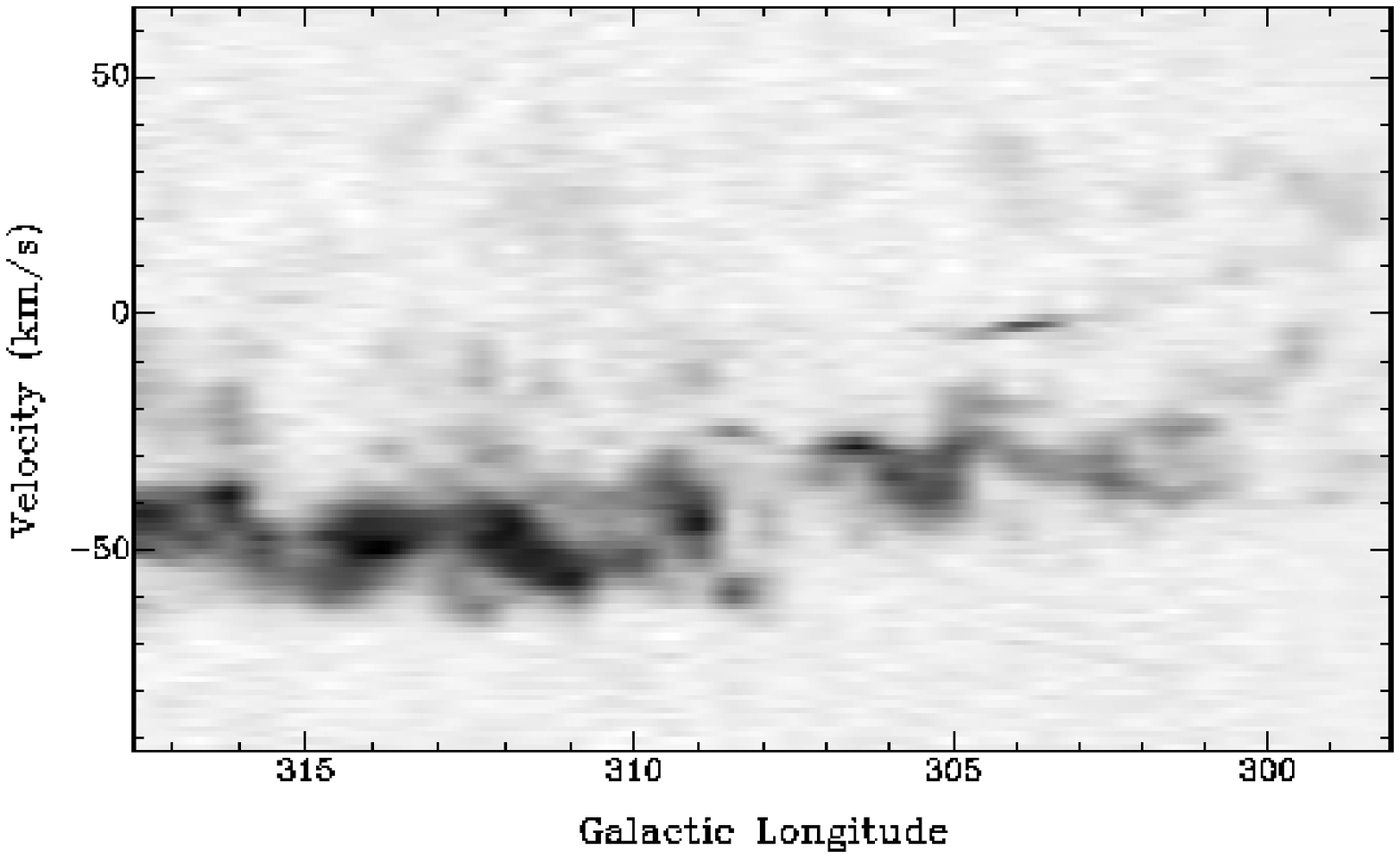}
\caption{Top: An $l-v$ diagram of the region ranging from $l$=298$^{\circ}$ to $l$=318$^{\circ}$.  The Sagittarius-Carina and Scutum-Crux spiral arms are labeled. Bottom:  An $l-v$ diagram of CO emission from the same region.  The CO brightness temperature ranges from -0.3 K (white) to 4.5 K (black).}
\end{figure}

\begin{figure}
\epsscale{1.7}
\plottwo{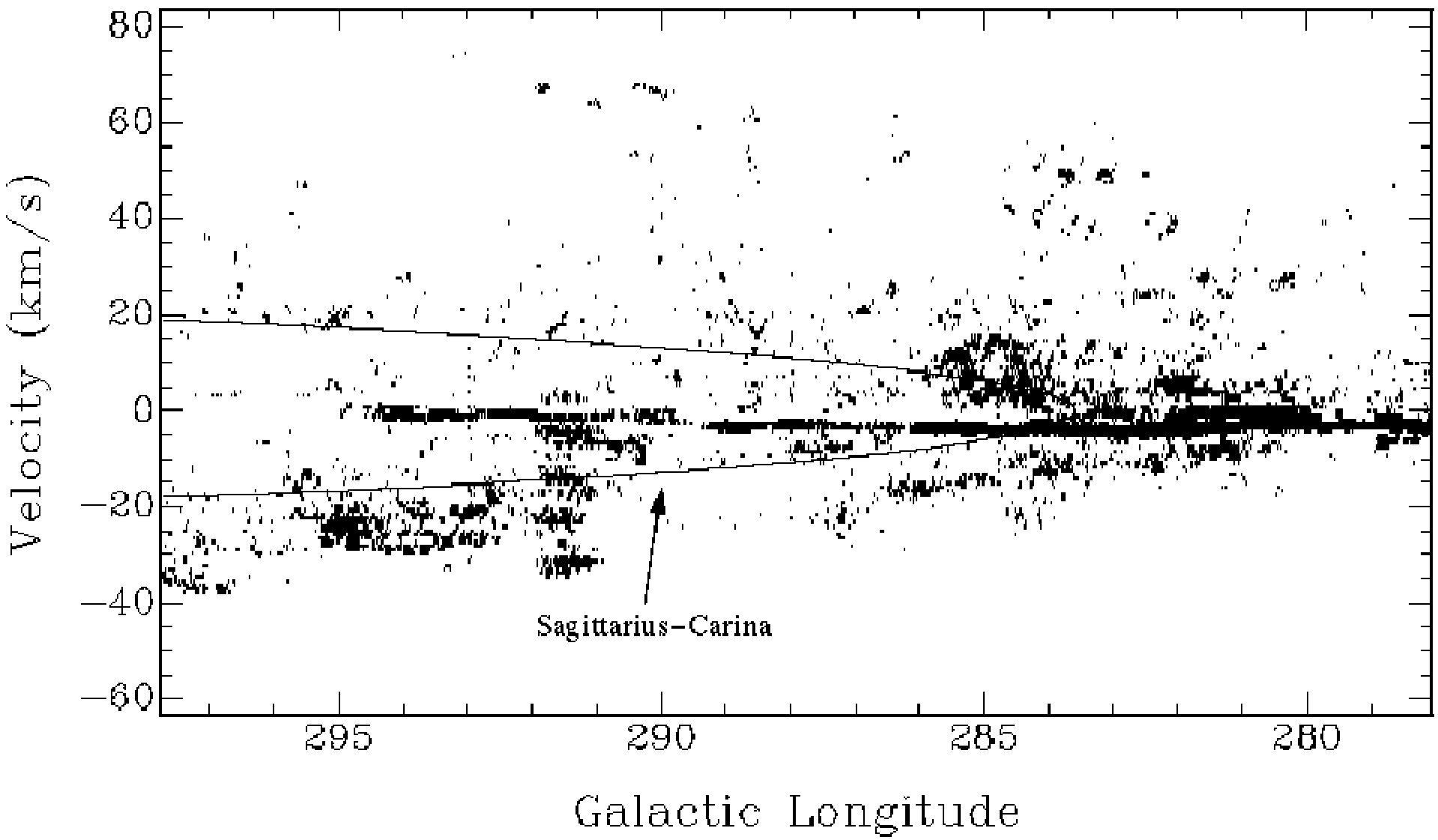}{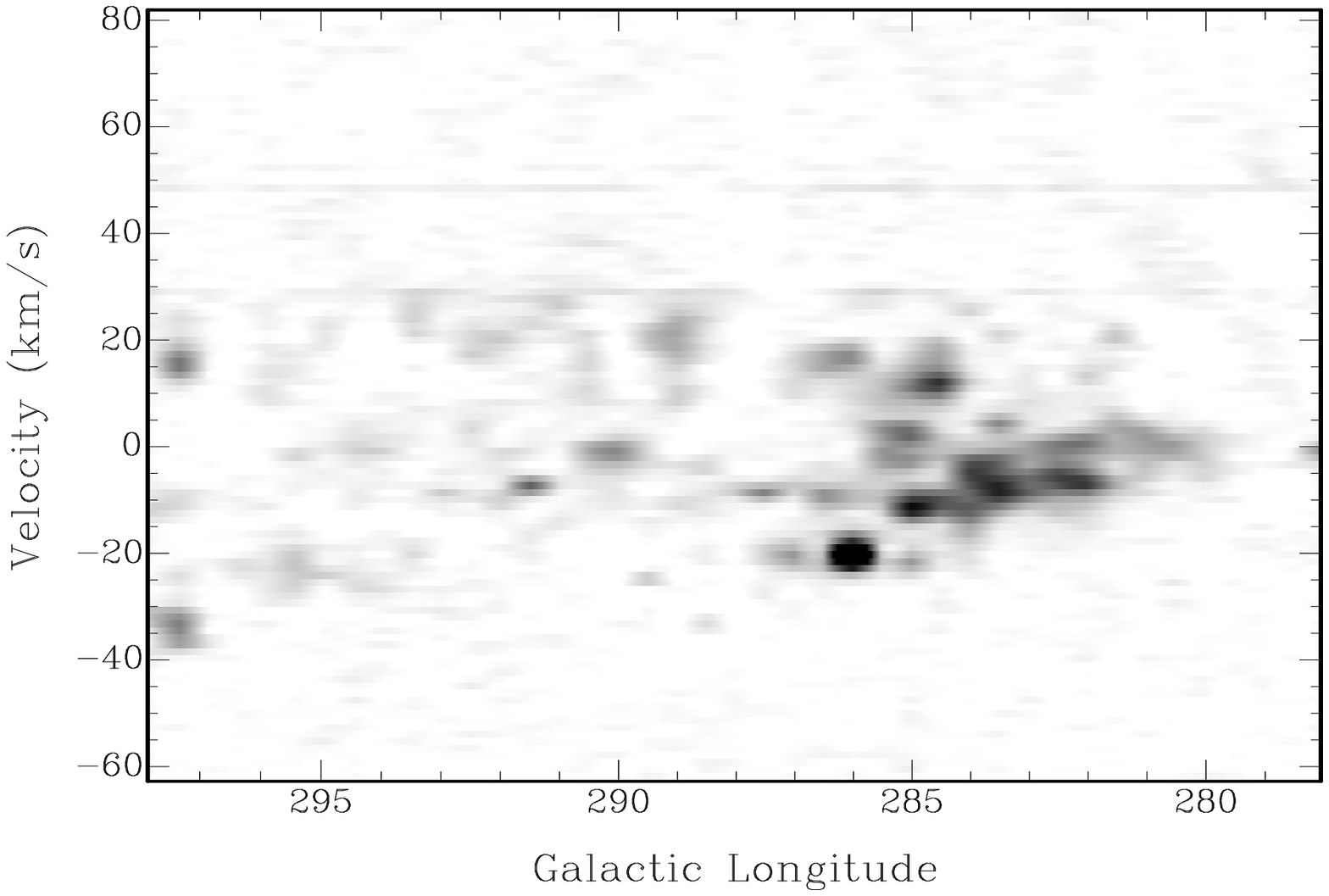}
\caption{Top: An $l-v$ diagram of the region ranging from $l$=278$^{\circ}$ to $l$=298$^{\circ}$.  The Sagittarius-Carina spiral arm is labeled.  Bottom:  An l-v diagram of CO emission from the same region.  The CO brightness temperature ranges from 0 K (white) to 2.0 K (black).}
\end{figure}

\begin{figure}
\epsscale{1.0}
\plotone{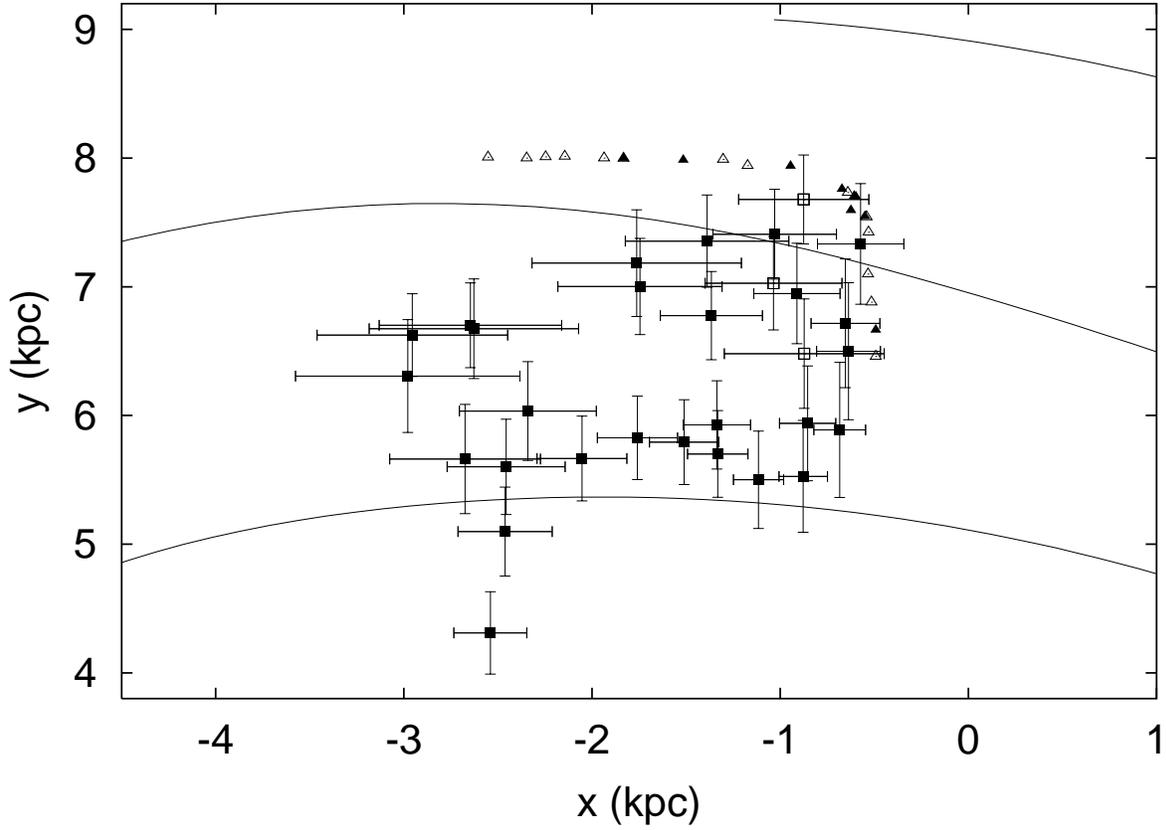}
\caption{HISA clouds from Table 2 plotted on the Galactic spiral arm pattern of \citet{2002astro.ph..7156C}.  The Scutum-Crux, Sagittarius-Carina, and local Orion-Cygnus spiral arms are drawn in increasing distance from the Galactic Center.  HISA clouds with a molecular counterpart are marked by filled squares and those without a molecular counterpart are marked by open squares.  Filled and open triangles represent HISA clouds, with and without a molecular counterpart, respectively, whose distances are given as upper limits in Table 2.   }
\end{figure}

 \begin{deluxetable}{ccc}
\tablewidth{0pc}
\tablecaption{Computed spin temperatures for each SGPS data subset.}
\tablehead{
\colhead{Central Longitude}  & \colhead{T$_s$($\tau$=$\infty$) (K)}  & \colhead{T$_s$($\tau$=1) (K)}}
\startdata

        260 & 52-65 & 31-48 \\
	265 & 38-55 & 15-38 \\
	270 & 41-52 & 17-37 \\
	275 & 51-67 & 31-51 \\
	280 & 52-72 & 26-53 \\
	285 & 42-67 & 23-52 \\
	290 & 39-57 & 16-39 \\
	295 & 51-68 & 30-49 \\
	300 & 47-63 & 19-42 \\
	305 & 48-68 & 21-47 \\
	310 & 59-74 & 35-54 \\
	315 & 42-67 & 19-42 \\
	320 & 38-54 & 16-32 \\
	325 & 34-57 & 15-38 \\
	330 & 45-66 & 23-46 \\
	335 & 49-64 & 24-45 \\
	340 & 50-69 & 26-49 \\
	345 & 51-66 & 22-45 \\ 

\enddata
\tablecomments{Spin temperature ranges are the 25-75\% distribution of HISA pixels in each SGPS subset.}
\end{deluxetable}

 \begin{deluxetable}{ccccccccccc}
\tablewidth{0pc}
\tablecaption{Computed properties for 70 HISA features}
\tablehead{
\colhead{\#}  & \colhead{l} & \colhead{b} & \colhead{V$_R$} & \colhead{D} & \colhead{d} & \colhead{T$_s$($\tau$=1)} & \colhead{T$_s$($\tau$=$\infty$)} & \colhead{$\Delta$V} & \colhead{N$_{HI}$} & \colhead{n$_{HI}$} \\
\colhead{} & \colhead{deg} & \colhead{deg} & \colhead{km s$^{-1}$} & \colhead{kpc} & \colhead{pc} & \colhead{K} & \colhead{K} & \colhead{km s$^{-1}$} & \colhead{10$^{19}$cm$^{-2}$} &  \colhead{cm$^{-3}$}}
\startdata

1 & 258.2 & -0.4 & +11.55 & 2.4$^{\ast}$ & 15 & 39 & 54  & 2.8 & 20 & 4.4 \\
2 & 261.4 & 0.7 & +3.31 & 3.3$^{\ast}$ & 40 & 35 & 50 & 2.3 & 15 & 1.2 \\
3 & 261.4 & 0.5 & +8.25 & 3.3$^{\ast}$ & 89 & 26 & 50  & 2.7 & 13 & 0.46 \\
4 & 263.8 & 0.8 & +7.43 & 4.5$^{\ast}$ & 110 & 24 & 42 & 2.9 & 13 & 0.39 \\
5 & 264.9 & 1.0 & +1.66 & 5.4$^{\ast}$ & 31 & 16 & 39 & 2.5 & 7.3 & 0.74 \\
6 & 265.5 & 0.9 & +14.85 & 6.1$^{\ast}$ & 17 & 6 & 24 & 3.6 & 3.9 & 0.76 \\
7 & 267.1 & 0.4 & +4.13  & 9.6$^{\ast}$ & 280 & 10 & 29 & 3.2 & 5.8 & 0.068 \\
8 & 267.3 & -0.3 & +6.61 & 10$^{\ast}$ & 240 & 11 & 31 &  3.2 & 6.4 & 0.088  \\
9 & 267.6 & 0.2 & +4.13 & 10$^{\ast}$ & 110 & 29 & 50 & 2.8 & 15 & 0.42  \\
10 & 268.2 & 0.9 & +14.85 & 10$^{\ast}$ & 79 & 13 & 34 & 3.8 & 9.0 & 0.37 \\
11 & 269.4 & -0.7 & +2.48 & 10$^{\ast}$ & 280 & 18 & 40 & 2.4 & 7.9 & 0.091 \\
12 & 270.3 & 0.6 & +0.01 & 10$^{\ast}$ & 400 & 16 & 41 & 2.6 & 7.6 & 0.061  \\
13 & 270.8 & 0.7 & +6.61 & 10$^{\ast}$ & 26 & 28 & 44 & 2.9 & 15 & 1.8  \\
14 & 273.1 & 0.7 & +3.31 & 9.0$^{\ast}$ & 55 & 35 & 52 & 2.5 & 16 & 0.94 \\
15 & 280.0 & -0.6 & -3.29 & 2.8$^{\ast}$ & 150 & 20 & 50 & 2.1 & 7.7 & 0.17  \\
16 & 280.3 & -0.7 & +27.20 & 6.9 & 12 & 23 & 40 & 2.9 & 12 & 3.3  \\
17 & 281.0 & 0.1 & +0.01 & 2.6$^{\ast}$ & 77 & 15 & 40 & 1.9 & 5.2 & 0.22  \\
18 & 282.0 & -0.7 & +5.78 & 2.4$^{\ast}$ & 25 & 21 & 41 & 2.2 & 8.4 & 1.1  \\
19 & 282.3 & -0.7 & -4.94 & 2.3$^{\ast}$ & 44 & 34 & 57 & 1.9 & 12 & 0.86  \\
20 & 282.5 & -0.6 & +23.92 & 7.0 & 49 & 16 & 41 & 3.7 & 11 & 1.9  \\
21 & 282.8 & -0.3 & -0.82 & 2.2$^{\ast}$ & 25 & 21 & 42 & 2.5 & 9.6 & 1.2 \\
22 & 284.5 & 0.0 & -4.12 & 2.0$^{\ast}$ & 84 & 26 & 45 & 2.3 & 11 & 0.42 \\
23 & 285.3 & -0.2 & +5.78 & 1.9$^{\ast}$ & 22 & 15 & 31 & 4.0 & 11 & 1.6 \\
24 & 285.4 & -0.3 & +11.55 & 1.9$^{\ast}$ & 12 & 27 & 55 & 3.9 & 19 & 5.4 \\
25 & 288.8 & 0.2 & -3.29 & 1.6$^{\ast}$ & 21 & 12 & 35 & 2.2 & 4.8 & 0.74  \\
26 & 291.4 & -0.2 & -6.59 & 1.4$^{\ast}$ & 7.3 & 17 & 39 & 2.5 & 7.7 & 3.4 \\
27 & 295.5 & 0.5 & -13.19 & 1.3$^{\ast}$ & 3.4 & 9 & 25 & 3.3 & 5.4 & 5.2 \\
28 & 300.7 & 0.0 & -7.42 & 1.1$^{\ast}$ & 6.7 & 8 & 26 & 3.3 & 4.8 & 3.4  \\
29 & 302.4 & -0.1 & -34.63 & 3.5 & 73 & 22 & 47 & 3.5 & 14 & 0.62  \\
30 & 304.2 & 0.0 & -33.80 & 3.2 & 61 & 29 & 55 & 2.8 & 15 & 0.78 \\	
31 & 304.8 & 0.3 & -35.45 & 3.2 & 110 & 12 & 38 & 3.8 & 8.3 & 0.24  \\
32 & 306.3 & 0.1 & -40.40 & 3.7 & 16 & 17 & 41 & 3.3 & 10 & 2.1  \\
33 & 306.7 & -0.4 & -28.85 & 2.2 & 71 & 25 & 54 & 4.5 & 21 & 0.94  \\
34 & 309.5 & -0.5 & -24.73 & 1.8 & 41 & 34 & 59 & 3.3 & 20 & 1.6  \\
35 & 310.6 & -0.1 & -31.33 & 2.3 & 46 & 37 & 59 & 3.4 & 23 & 1.6  \\
36 & 313.2 & 0.1 & -18.13 & 1.2 & 10 & 28 & 52 & 3.7 & 19 & 5.8  \\
37 & 316.5 & 0.5 & -47.82 & 3.4 & 12 & 20 & 42 & 3.1 & 11 & 3.1  \\
38 & 316.7 & -0.7 & -54.41 & 3.9 & 61 & 19 & 41 & 3.9 & 14 & 0.71  \\
39 & 316.7 & -0.4 & -22.26 & 1.5 & 17 & 41 & 55 & 2.6 & 19 & 3.7  \\
40 & 317.8 & -0.2 & -0.82 & 1.0$^{\ast}$ & 20 & 12 & 33 & 2.5 & 5.5 & 0.88  \\
41 & 319.7 & -0.6 & -55.24 & 3.8 & 30 & 21 & 45 & 4.1 & 16 & 1.7 \\
42 & 320.1 & -0.4 & -8.24 & 1.0$^{\ast}$ & 23 & 17 & 44 & 3.4 & 11 & 1.5  \\
43 & 321.6 & 0.0 & -32.98 & 2.2 & 44 & 17 & 32 & 2.0 & 6.2 & 0.45 \\
44 & 322.7 & -0.8 & +0.82 & 1.0$^{\ast}$ & 15 & 13 & 35 & 2.5 & 5.9 & 1.3 \\
45 & 323.3 & -0.1 & -11.54 & 1.0$^{\ast}$ & 12 & 12 & 32 & 4.1 & 9.0 & 2.4 \\
46 & 324.1 & -0.2 & -65.12 & 4.2 & 40 & 26 & 48 & 3.2 & 15 & 1.2 \\
47 & 324.1 & -0.4 & -52.75 & 3.5 & 12 & 9 & 38 & 4.0 & 6.6 & 1.7  \\
48 & 324.9 & 0.4 & -25.55 & 1.8 & 39 & 7 & 23 & 4.1 & 5.2 & 0.43  \\
49 & 325.4 & -0.0 & -8.25 & 1.1$^{\ast}$ & 23 & 9 & 25 & 4.3 & 7.0 & 0.99 \\
50 & 326.7 & 0.1 & -47.82 & 3.2 & 17 & 24 & 45 & 3.8 & 17 & 3.2 \\
51 & 328.8 & -0.4 & -79.15 & 4.9 & 26 & 24 & 44 & 3.9 & 17 & 2.2  \\
52 & 329.6 & 0.9 & -23.91 & 1.8 & 22 & 15 & 35 & 2.2 & 6.0 & 0.89 \\
53 & 330.0 & -0.1 & +3.30 & 1.1$^{\ast}$ & 17 & 20 & 38 & 2.3 & 8.4 & 1.6  \\
54 & 330.5 & -0.5 & -5.77 & 1.1$^{\ast}$ & 5.8 & 20 & 38 & 3.0 & 11 & 6.2  \\
55 & 330.8 & 0.2 & -44.52 & 3.1 & 35 & 35 & 55 & 2.5 & 16 & 1.5  \\
56 & 332.5 & 0.5 & -40.40 & 2.9 & 25 & 31 & 51 & 3.2 & 18 & 2.3  \\
57 & 333.7 & -0.9 & -13.19 & 1.2$^{\ast}$ & 6.3 & 10 & 31 & 2.8 & 5.1 & 2.6  \\
58 & 333.9 & -0.7 & -15.67 & 1.3 & 4.5 & 30 & 45 & 2.5 & 14 & 9.8  \\
59 & 334.5 & -0.4 & -43.70 & 3.1 & 35 & 27 & 50 & 3.2 & 16 & 1.5  \\
60 & 336.6 & -0.8 & -26.37 & 2.2 & 12 & 23 & 47 & 2.5 & 10 & 2.9  \\
61 & 339.2 & 0.4 & -3.30 & 1.5$^{\ast}$ & 31 & 34 & 58 & 22 & 15 & 2.2  \\
62 & 339.9 & -0.5 & -20.61 & 1.9 & 23 & 24 & 38 & 3.6 & 16 & 2.2  \\
63 & 340.6 & -0.5 & -37.93 & 3.2 & 45 & 30 & 50 & 3.2 & 18 & 1.3  \\
64 & 341.5 & 0.0 & -28.86 & 2.7 & 35 & 37 & 57 & 3.2 & 22 & 2.0  \\
65 & 342.3 & -0.2 & -20.61 & 2.1 & 27 & 28 & 52 & 2.6 & 13 & 1.6  \\
66 & 342.4 & -0.2 & +4.94 & 1.7$^{\ast}$ & 42 & 28 & 49 & 2.2 & 11 & 0.9  \\
67 & 343.5 & -0.8 & -32.16 & 3.1 & 30 & 31 & 52 & 4.2 & 24 & 2.6  \\
68 & 345.0 & -1.0 & +9.89 & 1.9$^{\ast}$ & 48 & 21 & 44 & 2.1 & 8.0 & 0.54 \\
69 & 345.3 & 0.1 & -24.74 & 2.7 & 16 & 31 & 47 & 2.0 & 11 & 2.2  \\
70 & 346.4 & -0.2 & +6.59 & 2.1$^{\ast}$ & 70 & 17 & 43 & 2.3 & 7.1 & 0.33 \\

\enddata
\tablecomments{For those clouds with negative radial velocities it is assumed that they are on the near side of the tangent point with an optical depth of 1.  An asterisk in Column 5 denotes an upper limit distance estimate.} 
\end{deluxetable}

\bibliography{references}
\end{document}